%% file: 0.main.tex
\documentclass[11pt]{article}

\usepackage[preprint]{acl}

\usepackage{times}
\usepackage{latexsym}

\usepackage{booktabs} 

\usepackage[T1]{fontenc}

\usepackage[utf8]{inputenc}

\usepackage{microtype}

\usepackage{inconsolata}

\usepackage{graphicx}
\usepackage{tabularx}

\usepackage{algorithm}
\usepackage[noend]{algpseudocode}

\usepackage{amsmath}
\usepackage{mathtools}
\usepackage{amssymb}
\usepackage{bm}

\usepackage{mathtools}
\usepackage{multirow}
\usepackage{subcaption}

\newcommand\doceq{\stackrel{\mathclap{\mbox{\tiny{doc}}}}{=}}

\DeclarePairedDelimiter{\card}{\lvert}{\rvert}

\NewDocumentCommand{\htag}{sm}{\texttt{<\IfBooleanT{#1}{/}#2>}}

\newcommand{\tok}[1]{\texttt{#1}}

\newcommand{\unk}{\tok{<unk>}}

\newcommand{\vq}{{\bm{q}}}
\newcommand{\vy}{{\bm{y}}}

\newcommand{\docs}{D}
\newcommand{\voc}{V}
\newcommand{\response}{\widehat{d}_\theta}

\newcommand{\docid}{\bm{v}_\text{docid}}
\newcommand{\doctxt}{\bm{v}_\text{text}}

\DeclareMathOperator*{\argmax}{arg\,max}
\DeclareMathOperator*{\argsortk}{arg\,sort^{\mathnormal{k}}}

\usepackage{listingsutf8}
\usepackage{csquotes}
\usepackage[safe]{tipa}
\usepackage{graphicx} 

\usepackage{threeparttable}

\usepackage{pgfplots}
\pgfplotsset{compat=1.18}
\definecolor{atomiccolor}{RGB}{0,158,115}
\definecolor{naivecolor}{RGB}{0,114,178}
\definecolor{semanticcolor}{RGB}{213,94,0}
\definecolor{naivelight}{RGB}{135,174,214}
\definecolor{naivemid}{RGB}{70,130,180}
\definecolor{naivedark}{RGB}{30,80,140}

\usepackage{tcolorbox}

\usepackage{array}
\usepackage{tabularx}
\newcolumntype{Y}{>{\raggedright\arraybackslash}X}

\usetikzlibrary{
  arrows.meta,
  positioning
}

\newcommand{\spmark}{%
  \raisebox{-0.20ex}{\rule{0.55em}{0.12ex}}%
}

\newcommand{\tenchildren}[4]{%
  \foreach \digit in {0,...,9}{%
    \pgfmathsetmacro{\childx}{(\digit-4.5)*0.48}%
    \ifnum\digit=#4\relax
      \node[selected choice]
        (#1-\digit)
        at (\childx,#3)
        {\digit};

      \draw[selected edge]
        (#2.south) -- (#1-\digit.north);
    \else
      \node[choice node]
        (#1-\digit)
        at (\childx,#3)
        {\digit};

      \draw[neutral edge]
        (#2.south) -- (#1-\digit.north);
    \fi
  }%
}

\newif\ifshowcomments
\showcommentstrue
\makeatletter
\ifacl@finalcopy\showcommentsfalse\fi
\makeatother

\title{\textsc{ReDSI}: Addressing the Reproducibility and Evaluation Consistency\\of Differentiable Search Indexing for Document Retrieval}

\author{
    Vivien Nicolas\textsuperscript{1,2} \qquad Hicham Randrianarivo\textsuperscript{1,3} \qquad Pascale Sébillot\textsuperscript{2} \qquad Caio Corro\textsuperscript{2}
    \\
    \textsuperscript{1}Artefact Research Center
    \\
    \textsuperscript{2}INSA Rennes, IRISA, CNRS, Université de Rennes
    \\
    \textsuperscript{3}MICS, CentraleSupélec, Université Paris-Saclay
    \\
    \small{
    \texttt{\{vivien.nicolas,hicham.randrianarivo\}@artefact.com}
    \quad
    \texttt{\{pascale.sebillot,caio.corro\}@irisa.fr}}
}

\begin{document}
\maketitle
\begin{abstract}
The differentiable search index (DSI) framework \citep{tay2022dsi} has become the \emph{de facto} baseline for generative retrieval.
However, DSI is hard to reproduce: no public implementation covers all three original document identifier types (atomic, naive, semantic), reported results vary widely, and the ubiquitous NQ320K dataset is built from Natural Questions through diverse and underspecified preprocessing.
We introduce \textsc{ReDSI}, the first open-source DSI implementation supporting all three identifier types, together with a parameterizable and well-documented NQ320K construction pipeline.

Experimentally,
we achieve results that are competitive with or stronger than previous DSI baselines.
Moreover, we conduct extensive experiments under model downscaling, covering retrieval effectiveness, parameter efficiency, training methods and decoding strategies,
opening novel directions for future research.
\end{abstract}

\input{1.Introduction}
\input{2.Background}
\input{3.NaturalQuestions}
\input{4.Experiments}

\input{5.Robustness}
\input{6.Construction}
\input{997.Conclusion}
\input{998.Limitations}

\bibliography{custom}

\clearpage
\appendix
\input{999.Appendices}

\end{document}

%% file: 1.Introduction.tex
\section{Introduction}

Generative retrieval (GR) has emerged as a new paradigm for information retrieval, in which documents relevant to a query are retrieved by generating their identifiers with an autoregressive language model \citep{cao2021entityretrieval,tay2022dsi}.\footnote{See the survey by \citet{li2025surveygr} and tutorial by \citet{tang2024tutorial}.}
Such systems require first defining how document identifiers are constructed, then fine-tuning a sequence-to-sequence model jointly for indexing and retrieval.
In practice, most systems use T5 \citep{raffel2020t5} as their backbone.

The seminal paper by \citet{tay2022dsi} known as \emph{differentiable search index} (DSI) has become a standard baseline for GR.
It proposes several methods for constructing document identifiers:
either using specific extra tokens in the decoder vocabulary (one token per document), called \emph{atomic} identifiers, or as strings of numbers using the tokenizer's original vocabulary.
For strings, they are either randomly assigned, called \emph{naive} identifiers or they follow a structured ``topic hierarchy'' that is induced in an unsupervised fashion, called \emph{semantic} identifiers.

However, we identify several limitations that prevent DSI from serving as a reliable baseline for future research, motivating the reproduction study presented in this paper.
First, no publicly available implementation of DSI includes all document identifier types introduced in the original paper, which hinders fair comparison.
Second, several papers report substantially lower results than the original paper using their own partial reimplementations \cite{chen-etal-2023-understanding,zhou2022ultron,bevilacqua2022autoregressive}.
Third, the most commonly used dataset for evaluation is Natural Questions \citep[NQ,][]{kwiatkowski2019nq} which was originally designed for a different task, namely question answering.
One must first transform NQ into a document retrieval dataset commonly known as NQ320K.
This requires several preprocessing decisions such as textual content extraction and deduplication strategy that can have a drastic impact on the resulting data, meaning that different authors use the same name NQ320K for substantially different datasets.
\emph{This makes it difficult to compare results reported by different authors}.

\begin{figure}[t]
\centering
\resizebox{0.9\linewidth}{!}{\input{pgf/model_scaling.pgf}}

\caption{MRR@10 versus number of parameters, including extra document token parameters for atomic identifiers.
Dotted, solid, and dashed lines denote atomic, naive, and semantic identifiers, respectively.
Results are reported for our \textsc{NQ320K-Title}/\textsc{NciNorm} dataset reconstruction.
}
\label{fig:model-scaling}
\end{figure}
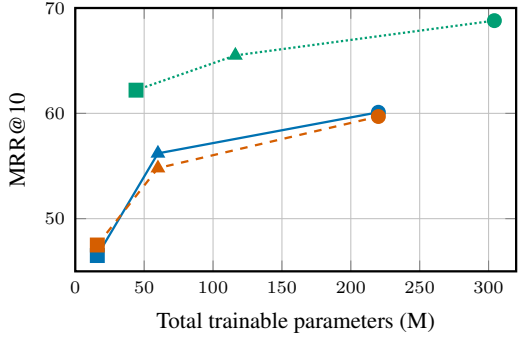

In this work, we introduce \textsc{ReDSI}, the first open-source implementation supporting all three original DSI identifier types under a common training and evaluation pipeline.
\textsc{ReDSI} also includes a parameterizable NQ320K construction pipeline.
This enables controlled analysis of dataset construction and modeling choices.

Previous work, including the original DSI paper, has primarily studied increasingly large T5 models \citep[\emph{inter alia}]{tay2022dsi,pradeep2023does}.
We argue that GR should be optimized for speed and therefore focus on robustness to downscaling the T5 backbone.
Our results show that atomic identifiers outperform naive and semantic identifiers at every evaluated model scale and retain substantially more effectiveness under downscaling the T5 backbone:
a 44M-parameter atomic model exceeds a 220M-parameter model using naive or semantic identifiers, see Figure~\ref{fig:model-scaling}.
Atomic identifiers are also faster to train.

Our contributions can be summarized as follows:
\begin{enumerate}
    \item We develop \textsc{ReDSI}, the first open-source DSI implementation supporting atomic, naive, and semantic identifiers under a common training and evaluation pipeline;\footnote{\url{https://github.com/vingtetun/ReDSI}}

    \item We formalize NQ320K construction and release a parameterizable pipeline
    that can produce both the datasets used in previous work and our newly proposed variants.\footnote{\url{https://huggingface.co/datasets/vingtetun/nq320k}}

    \item Using \textsc{ReDSI},
    we achieve results that are competitive with or stronger than previous DSI baselines;
    
    \item We conduct extensive experiments under model downscaling, covering retrieval effectiveness, parameter efficiency, training methods and decoding strategies.
\end{enumerate}

%% file: pgf/model_scaling.pgf
\begin{tikzpicture}
\begin{axis}[
    width=\linewidth,
    height=0.68\linewidth,
    xlabel={Total trainable parameters (M)},
    ylabel={MRR@10},
    xmin=0,
    xmax=320,
    ymin=45,
    ymax=70,
    xtick={0,50,100,150,200,250,300},
    grid=major,
    tick label style={font=\scriptsize},
    label style={font=\small},
    line width=0.9pt,
    mark size=2.5pt,
]

\addplot[
    atomiccolor,
    densely dotted,
    mark=none
]
coordinates {
    (44.1,62.2)
    (116.2,65.5)
    (304.2,68.8)
};

\addplot[
    naivecolor,
    solid,
    mark=none
]
coordinates {
    (16,46.5)
    (60,56.2)
    (220,60.1)
};

\addplot[
    semanticcolor,
    dashed,
    mark=none
]
coordinates {
    (16,47.5)
    (60,54.8)
    (220,59.7)
};

\addplot[
    only marks,
    mark=square*,
    mark options={solid, fill=atomiccolor, draw=atomiccolor}
]
coordinates {(44.1,62.2)};

\addplot[
    only marks,
    mark=square*,
    mark options={solid, fill=naivecolor, draw=naivecolor}
]
coordinates {(16,46.5)};

\addplot[
    only marks,
    mark=square*,
    mark options={solid, fill=semanticcolor, draw=semanticcolor}
]
coordinates {(16,47.5)};

\addplot[
    only marks,
    mark=triangle*,
    mark options={solid, fill=atomiccolor, draw=atomiccolor}
]
coordinates {(116.2,65.5)};

\addplot[
    only marks,
    mark=triangle*,
    mark options={solid, fill=naivecolor, draw=naivecolor}
]
coordinates {(60,56.2)};

\addplot[
    only marks,
    mark=triangle*,
    mark options={solid, fill=semanticcolor, draw=semanticcolor}
]
coordinates {(60,54.8)};

\addplot[
    only marks,
    mark=*,
    mark options={solid, fill=atomiccolor, draw=atomiccolor}
]
coordinates {(304.2,68.8)};

\addplot[
    only marks,
    mark=*,
    mark options={solid, fill=naivecolor, draw=naivecolor}
]
coordinates {(220,60.1)};

\addplot[
    only marks,
    mark=*,
    mark options={solid, fill=semanticcolor, draw=semanticcolor}
]
coordinates {(220,59.7)};

\end{axis}
\end{tikzpicture}

%% file: 2.Background.tex
\section{Background and Related Work}
\label{sec:background}

In this section, we first introduce our notation and summarize the differentiable search index (DSI) framework for document retrieval \citep{tay2022dsi}.
We then review subsequent works in generative retrieval and discuss the limitations of independently implemented DSI baselines that motivate our study.

\subsection{Differentiable Search Index}
\label{sec:dsi-background}

Let $p_\theta$ be an encoder-decoder model parameterized by $\theta \in \Theta$, with token vocabulary $\voc$.
Let $\docs$ be a document collection and $\doctxt: \docs \to \voc^*$ a map that associates to each document its textual content,
where $\voc^*$ is the Kleene closure of $\voc$.
In order to use $p_\theta$ to retrieve documents, we additionally need to define a map $\docid: \docs \to \voc^*$ that assigns a unique identifier to each document.

\paragraph{Document identifiers.}
\citet{tay2022dsi} proposed three identifier types.
\emph{Atomic} refers to the assignment of one dedicated output token to each document, written as \htag{doc\_0000}, \htag{doc\_0001}, etc.\footnote{Therefore the vocabulary is enriched by $\card{\docs}$ tokens.}
In this setting, we have $\card{\docid(d)} = 1, \forall d \in \docs$.

The other two types are based on tokenized decimal strings.
\emph{Naive} identifiers are arbitrary numerical sequences drawn from a predefined range,
\emph{i.e.}, numbers from $0$ to $b-1$ where $b$ is a budget hyperparameter such that $b \geq \card{\docs}$.
\emph{Semantic} identifiers instead encode paths through a hierarchy induced by clustering documents, see Appendix~\ref{app:semantic-algorithm}.

Following \citet{tay2022dsi}, for naive and semantic identifiers, the resulting decimal strings are tokenized using the model tokenizer,
meaning that decoder steps need not correspond to individual digits, and for semantic identifiers they do not necessarily align with semantic tree levels.
For example, \enquote{\tok{18510}} may become the following sequence of three tokens:
\enquote{\tok{18}}, \enquote{\tok{510}}, \htag{eos}.
These identifiers are multi-token, \emph{i.e.}, $\card{\docid(d)} \geq 2, \forall d \in \docs$.\footnote{Naive and semantic identifiers contain at least one decimal-string token followed by an end-of-sequence token.}

\paragraph{Retrieval.}
Given a user query $\vq \in \voc^*$, retrieval aims to search for the most relevant documents in $\docs$.
For any $d \in \docs$,
we assume that the model conditional probability $p_\theta(\docid(d) \mid \vq)$ can be used as a score of relevance, that is, the higher this probability, the more relevant the document is.
As such, retrieval reduces to searching for the ranked list of $k$ most probable documents:
\begin{align}\label{eq:retrieval}
    \response(\vq)
    =
    \argsortk_{d \in \docs}
    p_\theta(\docid(d) \mid \vq)\,,
\end{align}
where $\argsortk$ is the $\operatorname{arg\,sort}$ operator truncated at the $k$-th item.

The difficulty of computing $\response(\vq)$ depends on the identifier type.
For atomic identifiers, computing $\response(\vq)$ is trivial as all identifiers are composed of a single token, and hence it simplifies to computing $k$-best next tokens in the first position of the decoder.
When document identifiers are of arbitrary length, computing $\response(\vq)$ becomes intractable when the number of documents is large.
In practice, $\response(\vq)$ is approximated using beam search with width $k' \geq k$
\cite{lowerre1976harpy,graves2012seqtransduction}.

\paragraph{Training.}
To learn model $p_\theta$,
we rely on both $\docs$ and a training dataset $T \subset \voc^* \times \docs$,
where each couple $(\vq, d) \in T$ contains a user query $\vq$ and its associated most relevant document $d$.
The training objective is composed of an \emph{indexing} term that aims to maximize the log-likelihood of document identifiers given their textual contents,
and a \emph{retrieval} term that aims to maximize the log-likelihood of documents identifiers given user queries:
\begin{align}
\nonumber
    \widehat{\theta}\!=\!\argmax_{\theta \in \Theta}
    &
    \overbrace{\mathbb E_{d \sim \docs}\big[\,
        \log p_\theta(\docid(d) \mid \doctxt(d))
    \,\big]}^{\text{indexing}}
    \\
    &\null\hspace{-0.8cm}+
    \underbrace{
    \mathbb E_{(\vq, d) \sim T}\big[\,
        \log p_\theta(\docid(d) \mid \vq)
    \,\big]}_{\text{retrieval}}.
\label{eq:obj}
\end{align}
Maximization is done via stochastic gradient ascent \cite{bottou2003largescale}
with $\theta$ initialized from a pretrained model.

\subsection{Subsequent DSI Extensions}
\label{sec:gr-related-work}

The original DSI identifier schemes have not received equal attention. Much subsequent work has focused on designing or improving multi-token identifiers. Content-derived or learned identifiers include document substrings in SEAL \citep{bevilacqua2022autoregressive}, learned discrete codes in GenRet \citep{sun2023learning}, learnable $n$-grams in NOVO \citep{wang2023novo}, and permutation-invariant term sets in TSGen \citep{zhang2024generative}. NCI \citep{wang2022nci} instead modifies the decoder architecture to exploit hierarchical semantic identifiers. \citet{wu2024multivector} further relate multi-token generative retrieval to multi-vector dense retrieval \citep{khattab2020colbert}.

Although multiple studies investigate atomic identifiers for dynamic document collections \cite{mehta2023dsi++,kishore2023incdsi,huynh2025promptdsi,zhang2025dynamic},
atomic identifiers have received comparatively little dedicated attention in static-corpus settings.
\citet{nguyen2023grdense} relate atomic DSI to single-vector dense retrieval and propose to incorporate training techniques from the dense retrieval literature.

In this work, we argue that this imbalance is an important limitation of previous research.
We therefore compare the three original DSI identifier types on a fixed collection under a common implementation.
On NQ320K, the most commonly used DSI benchmark, we show that \emph{atomic identifiers outperform other types}, with the gap widening under model size downscaling.

Closer to our work, \citet{chen-etal-2023-understanding} analyze DSI's indexing and retrieval capabilities. \citet{pradeep2023does} examine corpus and model scaling across the original identifier types, including synthetic-query supervision, but do not release their code or processed data.
More recently, \citet{cai2025exploring} study multi-token identifiers, but again mainly focusing on upscaling the model size.
We instead implement all three original DSI identifier types in a common pipeline and explicitly examine their robustness when downscaling T5.

We refer readers to \citet{li2025surveygr} for a broader survey of generative retrieval.

\subsection{Prior DSI Baselines are not Comparable}
\label{sec:baselines_comparison}

\input{Tables/dsi_published_baselines}

Most generative retrieval works are evaluated on NQ320K, a dataset derived from Natural Questions \cite[NQ,][]{kwiatkowski2019nq}.
Because neither the original DSI implementation nor the generated NQ320K\footnote{The number \enquote{320K} refers to the number of queries in the NQ dataset, not the document collection size in NQ320K.} collection were publicly released, later studies reconstructed the baseline independently.
No published comparison covering all three original identifier types provides the complete combination of code, processed data and configurations required for exact reproduction.
Table~\ref{tab:published-dsi-nq} summarizes several baseline results reported in the literature.
\emph{We observe that the number of documents and baseline scores vary across reproductions, and that most works ignore atomic identifiers.}

These reconstructed baselines differ in their datasets, training, evaluation and pretrained backbones.
We address the dataset problem in Section~\ref{sec:nq320k}.
For training, the original DSI uses 32 times more samples for the Monte-Carlo estimation of the indexing expectation than for the retrieval one in Eq.~\eqref{eq:obj}, whereas later reimplementations often leave this information unspecified.
Other hyperparameters, such as the learning rate and number of optimization steps, differ across baselines or are incompletely reported.
Although most papers use T5-Base v1.0, some use Large, v1.1 or BART \cite{lewis2020bart}.
Finally, for evaluation, retrieval in Eq.~\eqref{eq:retrieval} may be approximated differently:
DSI uses beam width 40 and may return invalid sequences, whereas GenRet uses constrained decoding with beam width 100 \citep{sun2023learning}.
As such, Table~\ref{tab:published-dsi-nq} shows that reported baseline results in the literature are not comparable.

NCI has become an influential reconstruction because it releases preprocessing scripts, code, processed data, checkpoints and results \citep{wang2022nci,nci-nq-preprocessing}.
Although this contribution improves reproducibility, it propagates \emph{a particular NQ320K collection} and evaluates semantic identifiers exclusively.
Importantly, their NQ320K dataset differs from the original DSI paper in terms of document collection size.
Instead, we propose a dataset construction pipeline that reproduces several variants used in the literature,
including novel design possibilities, while supporting additional design choices, as described in Section~\ref{sec:nq320k}.

Note that differences in NQ320K construction even impact generative retrieval research beyond DSI,
\emph{e.g.}, \citet{mekonnen2025lightweight} compare their results with those of \citet{zhou2023genrrl}, although the two works use different dataset reconstructions.\footnote{``\emph{we deduplicate documents by
title}'' \cite[Sec.\ 5.1]{mekonnen2025lightweight} versus ``\emph{We eliminate duplicate
documents based on their URL}'' \cite[Sec.\ 5.1]{zhou2023genrrl}}

%% file: Tables/dsi_published_baselines.tex
\begin{table}[t]
\centering
\small
\setlength{\tabcolsep}{2.5pt}
\begin{tabular*}{\columnwidth}{@{\extracolsep{\fill}}lrcrrrc@{}}
\toprule
& & & \multicolumn{3}{c}{Hits@1} \\
\cmidrule(lr){4-6}
Reported by
& $\card{\docs}$
& T5
& At.
& Nv.
& Sm.
& \textsc{\textless/\textgreater}
\\
\midrule
\multicolumn{7}{@{}l}{\textbf{Complete identifier types coverage}} \\
\addlinespace[2pt]
Original DSI
& $\sim{}\!228\mathrm{K}$
& B
& 20.7
& 6.7
& 27.4
&
\\
\citet{pradeep2023does}
& $109.7\mathrm{K}$
& B\textsuperscript{+}
& 60.0
& 58.4
& 58.7
&
\\
\midrule
\multicolumn{7}{@{}l}{\textbf{Partial identifier types coverage}} \\
\addlinespace[2pt]
\citet{zhou2022ultron}
& $231.7\mathrm{K}$
& B
& 20.2
& \textendash
& 13.2
&
\\
\citet{bevilacqua2022autoregressive}
& $\sim{}\!200\mathrm{K}$
& L\textsuperscript{\(\circ\)}
& \textendash
& \textendash
& 25.0
&
\\
\citet{chen-etal-2023-understanding}
& $\sim{}\!200\mathrm{K}$
& B
& \textendash
& 0.1
& 2.1
& $\checkmark$
\\
\citet{wang2023novo}
& $109.7\mathrm{K}$
& B
& 49.4
& \textendash
& 38.2
&
\\
\citet{lee2023nonparametric}
& unk.
& L
& \textendash
& 12.5
& \textendash
&
\\
\citet{sun2023learning}
& $109.7\mathrm{K}$
& B
& \textendash
& \textendash
& 55.2
& $\checkmark$
\\
\citet{zhou2024roger}
& $231.7\mathrm{K}$
& B
& \textendash
& \textendash
& 27.4
&
\\
\citet{tang2024listwise}
& $228.0\mathrm{K}$
& B
& \textendash
& 22.1
& 27.9
& $\checkmark$
\\
\citet{zhang2024generative}
& $109.7\mathrm{K}$
& B
& \textendash
& \textendash
& 53.3
& $\checkmark$
\\
\midrule
\multicolumn{7}{@{}l}{\textbf{This work: ReDSI with different NQ320K variants}} \\
\textsc{title}/\textsc{NciNorm}
& $109.7\mathrm{K}$
& B
& 62.0
& 55.3
& 55.2
& $\checkmark$
\\
\textsc{title}/\textsc{HtmlNorm}
& $109.7\mathrm{K}$
& B
& 62.2
& 55.2
& 54.9
& $\checkmark$
\\
\textsc{pageid}/\textsc{NciNorm}
& $109.2\mathrm{K}$
& B
& 63.1
& 56.9
& 56.1
& $\checkmark$
\\
\textsc{pageid}/\textsc{HtmlNorm}
& $109.2\mathrm{K}$
& B
& 63.1
& 57.3
& 56.6
& $\checkmark$
\\
\bottomrule
\end{tabular*}
\caption{Hits@1 of published DSI baselines on independently NQ-derived collections for all identifier types.
T5 indicates the use of base (B) or large (L) version, with
 + for v1.1 instead of v1.0 and \(\circ\) for BART.
A checkmark indicates that released code contains an implementation of the reported DSI baseline.
For our work, we report several collection generation methods, see Section~\ref{sec:nq320k}.
%
}
\label{tab:published-dsi-nq}
\end{table}

%% file: 3.NaturalQuestions.tex
\section{Reconstructing NQ320K}
\label{sec:nq320k}

Natural Questions \citep[NQ,][]{kwiatkowski2019nq} is a question-answering dataset $Q$, where each example $(\vq,d,a)\in Q$ associates a user question $\vq$ with a Wikipedia page snapshot $d$ and an answer annotation $a$.
To repurpose NQ for document retrieval, a natural construction is to build a document collection \(\{d \mid \exists \vq,a:(\vq,d,a)\in Q\}\)
and query-document pairs $\{(\vq, d)\mid \exists a:(\vq,d,a)\in Q\}$ for training and evaluation.
This construction, however, is underspecified.

First, NQ stores a separate snapshot for every example, so the same Wikipedia page may appear in several revisions.\footnote{\url{https://en.wikipedia.org/wiki/Help:Page_history\#Linking_to_a_specific_revision_of_a_page}}
Constructing the document collection $\docs$ requires a deduplication rule that decides whether two snapshots should be merged as the same document, including across content revisions and title changes.

Second, each snapshot is a complete HTML page from which a textual input representation $\doctxt$ must be constructed.
We must extract the article's main textual content in a format that can then be used to build $\doctxt$.
We call this process \emph{serialization}.
As DSI-style models use only the first few dozen tokens of the serialized document, this process may have a significant impact on the indexing term in Equation~\eqref{eq:obj}.

\subsection{Document Deduplication}
\label{sec:nq-document-grouping}

\input{Tables/nq-document-collections-splits}

NQ stores one Wikipedia snapshot for each question, so the same underlying page may occur multiple times and under different revisions.
Constructing a retrieval collection therefore requires a comparison operator $\doceq$ to determine whether two snapshots should be considered as the same document.

We reimplement three methods from previous work:
\begin{itemize}
    \item $\doceq_\text{text4k}$ follows DSI by comparing the first 4000 characters of the serialized documents;
    \item $\doceq_\text{title}$ follows NCI by comparing tokenizer-normalized titles;\footnote{A page rename changes the title key, so revisions of the same Wikipedia page may be treated as distinct documents.}
    \item $\doceq_{\text{url}}$ follows Ultron by comparing revision-specific Wikipedia URLs.\footnote{A single-character edit is sufficient to create a new revision URL, causing two otherwise nearly identical snapshots of the same page to be treated as distinct documents.}
\end{itemize}
However,
all these comparison operators are sensitive to (possibly minor) content revision.
Comparing titles may even merge distinct pages.

We therefore propose to rely on the persistent Wikipedia \textsc{PageID} as a direct reference for page identity.
The resulting operator $\doceq_{\text{pageid}}$ considers two snapshots equivalent when they share the same PageID.\footnote{\url{https://www.mediawiki.org/wiki/Manual:Page_table\#page_id}}
PageIDs are generally preserved across revisions and article renames, making this rule better aligned with page-level retrieval.
Note that although we resolve $99.95\%$ of identifiers, we fall back on titles for the missing ones.\footnote{PageIDs are not provided directly in NQ, so we recover them by resolving the revision identifiers against a Wikipedia revision-history dump.}

Table~\ref{tab:nq-document-collections-splits} shows that the deduplication strategy can change the collection size by more than a factor of two.
It also changes query targets and identifier assignments used for training and evaluation.
Reproducing DSI's $\doceq_{\text{text4k}}$ rule is not straightforward.
The paper defines uniqueness from the first $4{,}000$ UTF-8 characters, but does not specify the source representation, normalization and truncation order, or whether \enquote{characters} denotes bytes or Unicode code points.
None of the interpretations we examined reproduces the reported 228K documents.
Our 202.2K-document reconstruction is instead close to the approximately 200K reported by \citet{bevilacqua2022autoregressive}.
Appendix~\ref{app:grouping-key-counts} reports deduplication variants and resulting collection sizes and Appendix~\ref{app:pageid-resolution} documents our PageID recovery.

\begin{table*}[t!]
\centering
\small
\input{Tables/data/example_both}
\caption{Two examples of the difference between serialization methods.
We show here the resulting mapping $\doctxt$ that returns the first 32 tokens using the T5 tokenizer,
where \texttt{\textless{}unk\textgreater{}} is the token representing unknown characters.}
\label{tab:ipa-tokenization-example}
\end{table*}

\subsection{Document Serialization}
\label{sec:nq-text-construction}

To simplify its use, NQ's authors provided a preprocessed version of the HTML content,
where the main article body is extracted from the raw HTML page and tokenized into words and tags \citep{nqtextutils}.
Nonetheless, it cannot be used directly as the extracted content contains a lot of extra information that one may want to remove, such as navigation content\footnote{\url{https://en.wikipedia.org/wiki/Wikipedia:Navigation_template}} and infoboxes.\footnote{\url{https://en.wikipedia.org/wiki/Help:Infobox}} 

\paragraph{Previous work (\textsc{NciNorm}).}
Existing DSI baselines use different methods to extract document content, often with insufficient pipeline details.
We therefore focus on the publicly available NCI preprocessing pipeline \citep{nci-nq-preprocessing}, which is a \emph{de facto} standard, and refer to it as \textsc{NciNorm}.

The document text is reconstructed from the preprocessed NQ data rather than from raw HTML.
Each document is formed by concatenating its title\footnote{For training records, the title is extracted from the first \htag{h1} element; for development records, it is read from the dataset's \texttt{document\_title} field. See Appendix~\ref{app:nq-formats}.} and first paragraph.\footnote{The content of the first \htag{p} element.}\textsuperscript{,}\footnote{The content after the first paragraph is only used to build semantic identifiers.}
A drawback of this approach is that it inherits artifacts from the NQ preprocessing pipeline, including incorrect word spacing and unwanted character substitutions,\footnote{Statistics on substitutions are reported in Appendix~\ref{app:nq-token-substitutions}.} which alter the resulting model tokenization.
This heuristic may also select the wrong content, \emph{e.g.}, the first \htag{p} element can occur inside an infobox.\footnote{Because infoboxes precede the article introduction, a \texttt{\textless{}p\textgreater{}} element within an infobox may be selected first.}

\paragraph{Proposed method (\textsc{HtmlNorm}).}
We introduce a configurable HTML-based reconstruction pipeline that reconstructs the text from the original HTML content, and can therefore bypass NQ preprocessing artifacts and preserve the original word tokenization.
Moreover, our pipeline is fully parameterizable, which allows future work to build different versions if necessary, \emph{i.e.}, with Wikipedia-specific elements to be retained, removed, or updated according to configurable rules.
The complete reconstruction, serialization, and cleanup procedure is described in Appendix~\ref{app:htmlnorm-details}.

In the \textsc{HtmlNorm} configuration used in this paper, we remove interface and navigation content, infoboxes, and other non-article artifacts, while transforming HTML elements such as lists and tables into text format.
We also remove leading parenthetical spans containing characters unsupported by the T5 tokenizer, typically pronunciation annotations in the International Phonetic Alphabet \citep{ipa1999}, because they produce several unknown tokens and can consume a substantial fraction of the fixed input budget.

Table~\ref{tab:ipa-tokenization-example} shows two examples of the textual content \(\doctxt\) produced by the two serialization methods.
\textsc{HtmlNorm} removes many unknown tokens.
Moreover, as \textsc{NciNorm} is derived from the preprocessed NQ data, it inherits incorrect word spacing that alters T5 tokenization and consumes additional input positions.
Details are given in Appendix~\ref{app:nq-tokenization-comparison}.

%% file: Tables/nq-document-collections-splits.tex
\begin{table}[t]
\centering
\small
\begin{tabular*}{\columnwidth}{@{\extracolsep{\fill}}lrrr@{}}
\toprule
\textbf{Collection} & \textbf{Train} & \textbf{Dev} & $|D|$ \\
\midrule
Original DSI dataset
& \textendash
& \textendash
& 228\,K \\
\midrule
\textsc{NQ320K-Text4K}
& 198.2\,K
& 7.3\,K
& 202.2\,K \\
\textsc{NQ320K-Title}
& 108.1\,K
& 6.9\,K
& 109.7\,K \\
\textsc{NQ320K-URL}
& 226.2\,K
& 7.4\,K
& 231.7\,K \\
\textsc{NQ320K-PageID}
& 107.6\,K
& 6.9\,K
& 109.2\,K \\
\bottomrule
\end{tabular*}
\caption{Number of documents produced by each deduplication rule in the training split, development split, and combined collection.}
\label{tab:nq-document-collections-splits}
\end{table}

%% file: Tables/data/example_both.tex
\begin{tabularx}{\textwidth}{@{}l>{\raggedright\arraybackslash\ttfamily}X@{}}
\toprule
\textsc{NciNorm}
& selena gomez selena marie gomez
( / s\unk li\unk n\unk\ \unk o\unk m \\
\textsc{HtmlNorm}
& selena gomez selena marie gomez
is an american singer and actress.
she began her career starring in the children \\
\midrule
\textsc{NciNorm}
& matt lanter matthew mackendree \unk\ matt
\char39\char39\ lanter ( born april 1, \\
\textsc{HtmlNorm}
& matt lanter matthew mackendree
\char34 matt\char34\ lanter (born april 1, 1983)
is an american \\
\bottomrule
\end{tabularx}

%% file: 4.Experiments.tex
\section{Experimental Setup}
\label{sec:experimental-setup}
\label{sec:reported-comparison}

\begin{table}[t!]
\centering
\small
\begin{tabular*}{\columnwidth}{@{\extracolsep{\fill}}lrrr@{}}
\toprule
&
\multicolumn{3}{c}{\textbf{MRR@10}} \\
\cmidrule(lr){2-4}
Ratio
& Atomic
& Naive
& Semantic \\
\midrule
$0{:}1$
& 56.35
& 50.03
& 49.92
\\
$1{:}2.8$ (natural)
& 69.42
& 60.07
& \textbf{59.10}
\\
$1{:}1$
& \textbf{69.76}
& \textbf{60.25}
& 58.33
\\
$4{:}1$
& 66.84
& 59.01
& 57.40
\\
$8{:}1$
& 64.15
& 58.95 
& 56.21
\\
$16{:}1$
& 61.44
& 58.49
& 55.35
\\
$32{:}1$
& 60.51
& 55.07
& 52.40
\\
$64{:}1$
& 59.29
& 50.88
& 48.87
\\
\bottomrule
\end{tabular*}
\caption{Sensitivity of T5-Base to the indexing-to-retrieval ratio on \textsc{NQ320K-Title}/\textsc{NciNorm}.}
\label{tab:indexing-retrieval-ratio-mrr}
\end{table}

To build $\doctxt$, we retain the first $\ell$ tokens given by the T5 tokenizer for each document.
For naive identifiers, we follow \citet{tay2022dsi} and set the assignment-space budget to $b=320\mathrm{K}$.

We evaluate on NQ development queries using Hits@1, Hits@10, and MRR@10.
Atomic identifiers are ranked directly from the first decoder-step distribution.
We use unconstrained beam search with width 40 for multi-token identifiers, followed by a filtering step against the set of valid identifiers.

We optimize all models with Adafactor \citep{shazeer2018adafactor}, using a constant learning rate, 1M updates and minibatches of size 128, with evaluation every 5K updates.
Checkpoints are selected by development MRR@10.

\paragraph{Hyperparameter search.}
We separately tune each identifier type over $\ell\in\{32,64\}$, learning rates $\{5\!\times\!10^{-4},10^{-3}\}$, and warmup durations $\{0,10\mathrm{K},100\mathrm{K}\}$.\footnote{We follow DSI choices for hyperparameter values.}
Complete search results and details are provided in Appendices~\ref{app:hyperparameters-search} and~\ref{app:reproduction-settings}.

\paragraph{Indexing-to-retrieval ratio.}
The expectations in Eq.\ \eqref{eq:obj} are estimated with different numbers of samples.
An indexing-to-retrieval ratio of $a{:}b$ means that for $a$ samples used to approximate the indexing term, there are $b$ samples used to approximate the retrieval term, no matter the minibatch size.
We call \enquote{natural ratio} the ratio that results from sampling all elements of $D$ and $T$ once, without replacement, in a full iteration over the training data.

Although this hyperparameter is widely overlooked (and even unreported) in the literature,
the results in Table~\ref{tab:indexing-retrieval-ratio-mrr} show that it has a drastic impact.
On average, the natural ratio $1{:}2.8$ works best, and we therefore use it instead of the $32{:}1$ ratio reported in DSI.
Extra results are given in Appendix~\ref{app:indexing-retrieval-ratio}.

\paragraph{Comparison to prior DSI baselines.}
Table~\ref{tab:published-dsi-nq} contextualizes \textsc{ReDSI} with a T5-Base backbone against published DSI baselines on NQ-derived collections.
Our results are comparable to the reproduction by \citet{pradeep2023does}, but our pipeline is publicly available.
When deduplicating by \textsc{Pageid} instead of \textsc{Title} as in NCI,
we obtain higher Hits@1 for both serialization methods (\textsc{NciNorm} and \textsc{HtmlNorm}).
By contrast, the two serialization methods yield small differences.
This may be because the model observes only the first $\ell$ tokens of each document and, under our minibatch sample composition, receives substantially more retrieval than indexing supervision, making the indexing term in Eq.~\eqref{eq:obj} comparatively less influential.
Nonetheless, \textsc{HtmlNorm} will be useful for future work that relies more on the textual content of documents.

%% file: 5.Robustness.tex
\section{Identifier Type Robustness under Downscaling}
\label{sec:identifier-analysis}

\begin{table*}[t]
\centering
\small
\setlength{\tabcolsep}{3pt}
\renewcommand{\arraystretch}{0.9}
\input{Tables/nq320k_identifier_comparison}
\caption{Comparison across document-text constructions, deduplication rules, identifier types, and T5 model scales. Configuration denotes the document-text construction and deduplication rule; all values are percentages. Bold marks the highest MRR@10 among the four reconstruction configurations for each identifier type and model scale.}
\label{tab:nq320k_identifier_comparison}
\end{table*}

We cross two deduplication rules, $\doceq_{\textsc{title}}$ and $\doceq_{\textsc{pageid}}$, with two document representations, $\textsc{NciNorm}$ and $\textsc{HtmlNorm}$.
This $2\times2$ design varies each reconstruction choice while holding the other fixed.
The \textsc{NQ320K-Title}/\textsc{NciNorm} configuration follows the influential NCI reconstruction.
Replacing \textsc{NciNorm} with \textsc{HtmlNorm} tests HTML-aware surface reconstruction, whereas replacing
$\doceq_{\text{title}}$ with $\doceq_{\text{pageid}}$ tests persistent page identity rather than normalized titles.
We compare three T5 backbone scales:
Efficient-Tiny (16M parameters),
Small (60M)
and Base (220M).\footnote{At the smaller scales, we reuse the identifier-specific hyperparameters selected for T5-Base without retuning.}

Table~\ref{tab:nq320k_identifier_comparison} reports results across identifier types and model scales.
Supplementary results about variance across training runs are given in Appendix~\ref{app:nq320k-construction-variability}.

\paragraph{Atomic identifiers always outperform other types.}
Atomic identifiers introduce $\card{\docs}$ document-specific output embeddings.
The resulting atomic models contain 44M, 116M, and 304M trainable parameters for T5-Efficient-Tiny, T5-Small, and T5-Base, respectively.
Parameter count also understates the inference advantage of atomic identifiers, which require one decoder step rather than autoregressive generation.

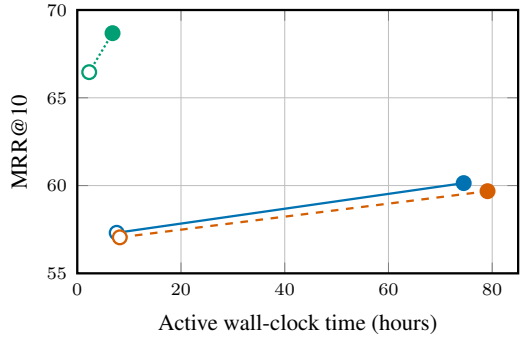
\begin{figure}[t]
\centering
\resizebox{0.9\linewidth}{!}{\input{pgf/training_dynamics.pgf}}

\caption{%
MRR@10 on \textsc{NQ320K-Title}/\textsc{NciNorm} versus wall-clock time for T5-Base.
Hollow markers show the first checkpoint reaching $95\%$ of peak MRR@10, and filled markers show the final selected checkpoint.
Dotted, solid, and dashed lines denote atomic, naive, and semantic identifiers.
}
\label{fig:model-scaling-and-training}
\end{figure}

Figure~\ref{fig:model-scaling} shows that atomic identifiers are substantially more robust to backbone downscaling.
From T5-Base to T5-Efficient-Tiny, atomic MRR@10 decreases by $6.6$ points, compared with $13.6$ for naive and $12.2$ for semantic identifiers.
Moreover, the 44M-parameter atomic model reaches $62.2$ MRR@10, outperforming the 220M-parameter T5-Base models with naive ($60.1$) and semantic ($59.7$) identifiers.
Atomic identifiers therefore retain effectiveness substantially better under downscaling and provide the strongest effectiveness--parameter trade-off.

\paragraph{Atomic identifiers enable faster training.}
As shown in Figure~\ref{fig:model-scaling-and-training}, for T5-Base,
training with atomic identifiers reaches its peak MRR@10 after 6.8 hours ($\sim\!65\mathrm{K}$ updates), whereas naive and semantic identifiers continue improving beyond 74.5 hours ($\sim\!900\mathrm{K}$ updates).
Atomic identifiers thus combine higher effectiveness with substantially faster training.

\begin{figure}[t]
\centering
\resizebox{\columnwidth}{!}{%
\input{pgf/naive_semantic_invalid_rates.pgf}
}
\caption{Post-filtered MRR@10 and Invalid@40 across model scales on \textsc{NQ320K-Title}/\textsc{NciNorm}. Solid and dashed lines denote naive and semantic identifiers, while square, triangle, and circle markers denote T5-Efficient-Tiny, T5-Small, and T5-Base.}
\label{fig:identifier-validity}
\end{figure}
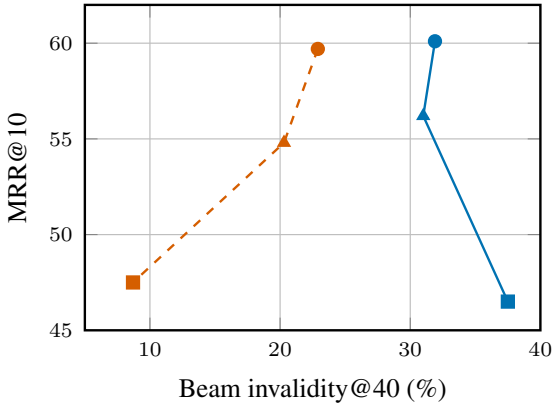

\paragraph{Semantic identifiers are more robust to invalid generation.}
Models with multi-token identifiers may generate invalid identifiers.
In Figure~\ref{fig:identifier-validity},
we compare \emph{Invalid@40}, the mean proportion of invalid hypotheses in the unconstrained 40-best beam, with MRR@10.
We observe that semantic identifiers produce fewer invalid hypotheses than naive identifiers at every scale.

For naive identifiers the proportion of invalid identifiers increases with the smallest backbone,
whereas it drastically decreases for semantic identifiers.
Moreover, for this smallest model the MRR@10 is higher for semantic identifiers ($47.5$ MRR@10 compared with $46.5$).
This suggests that semantic structure may become more beneficial at smaller model scales, opening up avenues for future research on downscaling DSI models.
Additional comparisons are given in Appendix~\ref{app:invalid-40}.

\paragraph{Downscaling does not require constrained decoding.}
A common approach to guarantee valid generated identifiers is to rely on constrained decoding \citep{cao2021entityretrieval}.\footnote{See Appendix \ref{app:additional-decoding-results} for further details.}
Table~\ref{tab:decoding-by-model-scale} compares unconstrained decoding, unconstrained decoding with post-filtering (\emph{i.e.}, removing invalid identifiers) and constrained decoding.
Constrained decoding provides no consistent improvement over simple post-filtering, even when downscaling the model. Thus, in our setting, smaller T5 backbones do not require complex decoding to retain retrieval effectiveness.

\input{Tables/constrained-decoding}

%% file: Tables/nq320k_identifier_comparison.tex
\begin{tabular*}{\textwidth}{@{\extracolsep{\fill}}ll*{9}{r}@{}}
\toprule
&
&
\multicolumn{3}{c}{\textbf{Atomic}} &
\multicolumn{3}{c}{\textbf{Naive}} &
\multicolumn{3}{c}{\textbf{Semantic}} \\
\cmidrule(lr){3-5}
\cmidrule(lr){6-8}
\cmidrule(lr){9-11}
Configuration
& Model
& H@1 & H@10 & MRR@10
& H@1 & H@10 & MRR@10
& H@1 & H@10 & MRR@10 \\
\midrule

\multirow{3}{*}{\textsc{NQ320K-Title}/\textsc{NciNorm}}
& Base
& 62.0 & 81.2 & 68.8
& 55.3 & 69.4 & 60.1
& 55.2 & 68.4 & 59.7 \\

& Small
& 58.6 & 78.3 & 65.5
& 51.5 & 64.5 & 56.2
& 50.1 & 63.9 & 54.8 \\

& Eff.-Tiny
& 55.6 & 74.9 & 62.2
& 42.4 & 53.8 & 46.5
& 42.9 & 56.1 & 47.5 \\

\addlinespace[2pt]

\multirow{3}{*}{\textsc{NQ320K-PageID}/\textsc{NciNorm}}
& Base
& 63.1 & 82.2 & 69.8
& 56.9 & 70.2 & 61.6
& 56.1 & 70.1 & 60.9 \\

& Small
& 59.8 & 79.0 & 66.5
& 52.3 & 65.6 & 57.0
& 51.4 & 65.3 & \textbf{56.1} \\

& Eff.-Tiny
& 56.4 & 75.7 & 63.2
& 43.4 & 55.6 & \textbf{47.8}
& 44.1 & 57.5 & \textbf{48.8} \\

\midrule

\multirow{3}{*}{\textsc{NQ320K-Title}/\textsc{HtmlNorm}}
& Base
& 62.2 & 81.2 & 68.8
& 56.2 & 69.8 & 61.0
& 54.9 & 69.4 & 60.0 \\

& Small
& 59.4 & 78.4 & 65.9
& 51.6 & 64.8 & 56.2
& 50.4 & 64.6 & 55.2 \\

& Eff.-Tiny
& 55.8 & 75.2 & 62.5
& 42.4 & 55.4 & 47.0
& 43.3 & 56.6 & 47.8 \\

\addlinespace[2pt]

\multirow{3}{*}{\textsc{NQ320K-PageID}/\textsc{HtmlNorm}}
& Base
& 63.1 & 82.6 & \textbf{70.0}
& 57.3 & 70.5 & \textbf{62.0}
& 56.6 & 69.8 & \textbf{61.1} \\

& Small
& 60.0 & 79.6 & \textbf{66.9}
& 52.4 & 65.5 & \textbf{57.1}
& 51.1 & 65.0 & 55.9 \\

& Eff.-Tiny
& 56.8 & 75.8 & \textbf{63.4}
& 42.6 & 55.1 & 47.0
& 44.2 & 57.6 & 48.7 \\

\bottomrule
\end{tabular*}

%% file: pgf/training_dynamics.pgf
\begin{tikzpicture}
\begin{axis}[
    width=\linewidth,
    height=0.68\linewidth,
    xlabel={Active wall-clock time (hours)},
    ylabel={MRR@10},
    xmin=0,
    xmax=85,
    ymin=55,
    ymax=70,
    xtick={0,20,40,60,80},
    ytick={55,60,65,70},
    grid=major,
    tick label style={font=\scriptsize},
    label style={font=\small},
    line width=0.9pt,
    mark size=2.7pt,
]

\addplot[
    atomiccolor,
    densely dotted,
    mark=none
]
coordinates {
    (2.3,66.46)
    (6.8,68.68)
};

\addplot[
    naivecolor,
    solid,
    mark=none
]
coordinates {
    (7.6,57.31)
    (74.5,60.14)
};

\addplot[
    semanticcolor,
    dashed,
    mark=none
]
coordinates {
    (8.2,57.05)
    (79.1,59.68)
};

\addplot[
    only marks,
    mark=*,
    mark options={
        solid,
        fill=white,
        draw=atomiccolor,
        line width=0.9pt
    }
]
coordinates {(2.3,66.46)};

\addplot[
    only marks,
    mark=*,
    mark options={
        solid,
        fill=white,
        draw=naivecolor,
        line width=0.9pt
    }
]
coordinates {(7.6,57.31)};

\addplot[
    only marks,
    mark=*,
    mark options={
        solid,
        fill=white,
        draw=semanticcolor,
        line width=0.9pt
    }
]
coordinates {(8.2,57.05)};

\addplot[
    only marks,
    mark=*,
    mark options={
        solid,
        fill=atomiccolor,
        draw=atomiccolor
    }
]
coordinates {(6.8,68.68)};

\addplot[
    only marks,
    mark=*,
    mark options={
        solid,
        fill=naivecolor,
        draw=naivecolor
    }
]
coordinates {(74.5,60.14)};

\addplot[
    only marks,
    mark=*,
    mark options={
        solid,
        fill=semanticcolor,
        draw=semanticcolor
    }
]
coordinates {(79.1,59.68)};

\end{axis}
\end{tikzpicture}

%% file: pgf/naive_semantic_invalid_rates.pgf
\begin{tikzpicture}
\begin{axis}[
    scale only axis,
    width=0.70\linewidth,
    height=0.50\linewidth,
    xlabel={Beam invalidity@40 (\%)},
    ylabel={MRR@10},
    xmin=5,
    xmax=40,
    ymin=45,
    ymax=62,
    grid=major,
    tick label style={font=\scriptsize},
    label style={font=\small},
    line width=0.8pt,
]

\addplot[
    naivecolor,
    solid,
    mark=none
]
coordinates {
    (31.9,60.1)
    (31.0,56.2)
    (37.5,46.5)
};

\addplot[
    semanticcolor,
    dashed,
    mark=none
]
coordinates {
    (22.9,59.7)
    (20.3,54.8)
    (8.7,47.5)
};

\addplot[
    only marks,
    mark=*,
    mark options={solid, fill=naivecolor, draw=naivecolor}
]
coordinates {(31.9,60.1)};

\addplot[
    only marks,
    mark=*,
    mark options={solid, fill=semanticcolor, draw=semanticcolor}
]
coordinates {(22.9,59.7)};

\addplot[
    only marks,
    mark=triangle*,
    mark options={solid, fill=naivecolor, draw=naivecolor}
]
coordinates {(31.0,56.2)};

\addplot[
    only marks,
    mark=triangle*,
    mark options={solid, fill=semanticcolor, draw=semanticcolor}
]
coordinates {(20.3,54.8)};

\addplot[
    only marks,
    mark=square*,
    mark options={solid, fill=naivecolor, draw=naivecolor}
]
coordinates {(37.5,46.5)};

\addplot[
    only marks,
    mark=square*,
    mark options={solid, fill=semanticcolor, draw=semanticcolor}
]
coordinates {(8.7,47.5)};

\end{axis}
\end{tikzpicture}

%% file: Tables/constrained-decoding.tex
\begin{table}[t]
\centering
\small
\setlength{\tabcolsep}{2pt}
\begin{tabular*}{\columnwidth}{@{\extracolsep{\fill}}llrrrrr@{}}
\toprule
Identifier
& Model
& Unc.
& Post.
& Trie
& $\Delta_{\mathrm{post}}$
& $\Delta_{\mathrm{trie}}$ \\
\midrule
\multirow{3}{*}{Naive}
& T5-Base
& 59.53
& 59.98
& 59.99
& $+0.45$
& $+0.01$ \\
& T5-Small
& 55.80
& 56.25
& 56.26
& $+0.45$
& $+0.01$ \\
& T5-Eff.-Tiny
& 45.73
& 46.51
& 46.53
& $+0.78$
& $+0.02$ \\
\midrule
\multirow{3}{*}{Semantic}
& T5-Base
& 59.46
& 59.61
& 59.21
& $+0.15$
& $-0.40$ \\
& T5-Small
& 54.69
& 54.83
& 54.23
& $+0.14$
& $-0.60$ \\
& T5-Eff.-Tiny
& 47.42
& 47.50
& 47.52
& $+0.08$
& $+0.02$ \\
\bottomrule
\end{tabular*}
\caption{MRR@10 across model scales under unconstrained (\emph{Unc.}), post-filtered (\emph{Post.}) and trie-constrained (\emph{Trie}) decoding on \textsc{NQ320K-Title}/\textsc{NciNorm}. $\Delta_{\mathrm{post}}$ is post-filtered minus unconstrained MRR@10, and $\Delta_{\mathrm{trie}}$ is trie-constrained minus post-filtered MRR@10.}
\label{tab:decoding-by-model-scale}
\end{table}

%% file: 6.Construction.tex
\section{Qualitative Analysis of NQ320K}
\label{sec:construction-results}

In this section, we focus on the NQ320K dataset construction pipeline analysis.

\paragraph{Document deduplication.}
$\doceq_{\text{title}}$ fragments 550 resolved Wikipedia pages and creates cross-page collisions involving 93 PageIDs, affecting $1.62\%$ of training queries and $2.54\%$ of development queries.
$\doceq_{\text{pageid}}$ deduplication associates 1,054 additional training queries with development-target documents, increasing supervision coverage from $75.3\%$ to $76.4\%$.
Table~\ref{tab:title-grouping-examples} illustrates title-induced page fragmentation and cross-page collisions.

\paragraph{Document-text construction.}
Within the first 32 T5 tokens, \textsc{NciNorm} and \textsc{HtmlNorm} differ for $92.6\%$ of documents, while the proportion containing at least one \unk\ falls from $18.7\%$ to $1.6\%$.

\input{Tables/title-grouping-examples}

%% file: Tables/title-grouping-examples.tex
\begin{table}[t]
\centering
\small
\begin{threeparttable}
\begin{tabularx}{\columnwidth}{@{}lYY@{}}
\toprule
& Document A & Document B \\
\midrule
\multicolumn{3}{@{}l}{\textbf{Fragmentation: one PageID, multiple titles}} \\
\addlinespace[2pt]
Title
& \texttt{Elsa (Disney)}
& \texttt{Elsa (Frozen)} \\
PageID
& \multicolumn{2}{c}{\texttt{41312390}} \\
\midrule
\multicolumn{3}{@{}l}{\textbf{Collision: one title key, multiple PageIDs}} \\
\addlinespace[2pt]
Title
& \texttt{It Takes a Village}
& \texttt{It takes a village} \\
PageID
& \texttt{2308910}
& \texttt{12185842} \\
\bottomrule
\end{tabularx}
\begin{tablenotes}[flushleft]
\footnotesize
\item[] Wikipedia pages can be accessed using
\url{https://en.wikipedia.org/?curid=PAGEID}.
\end{tablenotes}
\caption{Examples of fragmentation and cross-page collision under
\(\doceq_{\text{title}}\).
}
\label{tab:title-grouping-examples}
\end{threeparttable}
\end{table}

%% file: 997.Conclusion.tex
\section{Conclusion}
\label{sec:conclusion}

We release \textsc{ReDSI}, an open-source DSI implementation supporting all original identifier types, together with a parameterizable NQ320K construction pipeline.
By separating document deduplication from serialization, we show that both choices affect retrieval effectiveness: \textsc{NQ320K-PageID} yields larger and more consistent gains over \textsc{NQ320K-Title}, while \textsc{HtmlNorm} provides modest improvements over \textsc{NciNorm}.

Moreover, we conduct extensive experiments showing the robustness of the DSI framework when downscaling the T5 backbone.
Atomic identifiers achieve the strongest effectiveness, degrade least under downscaling, and train substantially faster.

Together, \textsc{ReDSI} and the named NQ320K variants provide a reproducible basis for evaluating future DSI-style systems.
Our \textsc{HtmlNorm} variant will be especially useful for future work that relies more on the textual content of documents.

%% file: 998.Limitations.tex
\section*{Limitations}

Our controlled experiments are restricted to English NQ-derived collections and T5-family backbones ranging from T5-Efficient-Tiny to T5-Base.
The relative behavior of identifier types may differ on other corpora, tokenizers, architectures, or substantially larger and dynamic collections.
In particular, atomic identifiers add one output vector per document, so their favorable effectiveness--parameter trade-off on a 109.7K-document collection does not establish their scalability to millions of documents.
Moreover, the smaller backbones reuse hyperparameters selected on T5-Base; this deliberately measures robustness under a fixed training configuration rather than the best attainable performance at each scale.
Unfortunately, extensive grid search is costly in terms of computational resources.

%% file: 999.Appendices.tex
\input{Appendices/semantic-algorithm}
\input{Appendices/grouping-keys}
\input{Appendices/htmlnorm-details}
\input{Appendices/natural-questions}
\input{Appendices/indexing-retrieval-ratio}
\input{Appendices/hyperparameters-search}
\input{Appendices/reproduction-settings}
\input{Appendices/nq320k-construction-variability}
\input{Appendices/invalid-40}
\input{Appendices/additional-decoding-results}

%% file: Appendices/semantic-algorithm.tex
\section{Semantic Identifiers}
\label{app:semantic-algorithm}

To assign semantic identifiers to documents, we follow DSI's original decimal-tree construction \citep{tay2022dsi}. Because our dataset-construction pipeline changes both the document collection and its textual contents (Section~\ref{sec:nq320k}), we cannot reuse the semantic identifiers released by \citet{nci-nq-preprocessing}.

Documents are embedded with \texttt{google/bert\_uncased\_L-8\_H-512\_A-8} \cite{devlin-etal-2019-bert} from their text, truncated to 512 tokens. We then recursively partition the document embeddings using scikit-learn's \texttt{KMeans} with default parameters, creating 10 clusters at each internal node. Recursion stops once a cluster contains 100 or less documents. The sequence of cluster labels encountered along the root-to-leaf path forms a shared decimal prefix, after which documents within the terminal cluster receive consecutive local indices to ensure uniqueness.
Figure~\ref{fig:semantic-identifier-tokenization} illustrates both the hierarchical construction of the identifier and its subsequent tokenization by T5.

We also illustrate naive tokenization in Figure~\ref{fig:naive-identifier-tokenization} to emphasize the difference.

\input{Appendices/semantic-clustering}
\input{Appendices/naive-identifier}


%% file: Appendices/semantic-clustering.tex
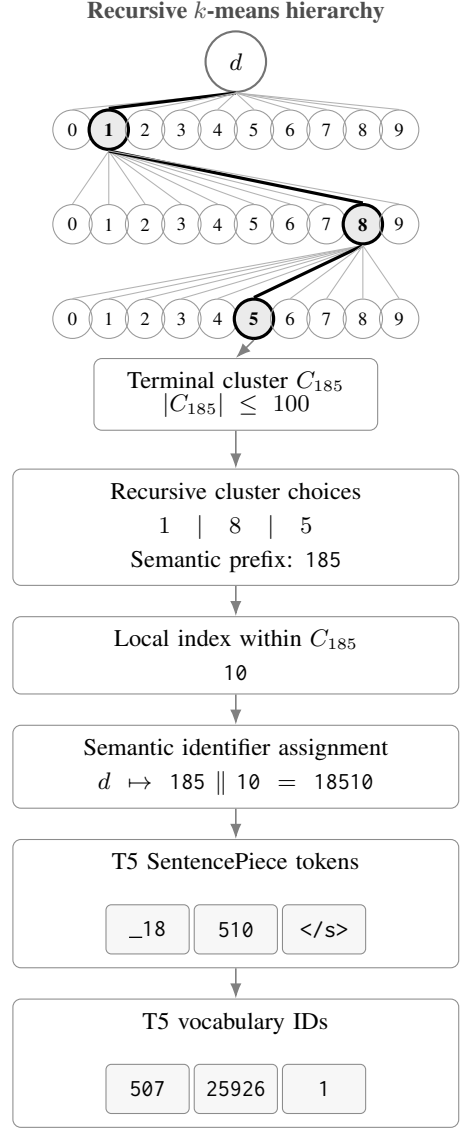
\begin{figure}[t!]
\centering
\begin{tikzpicture}[
    font=\small,
    >=Latex,
    root node/.style={
        circle,
        draw=black!55,
        thick,
        minimum size=8mm,
        inner sep=0pt
    },
    choice node/.style={
        circle,
        draw=black!40,
        minimum size=5.2mm,
        inner sep=0pt,
        font=\scriptsize
    },
    selected choice/.style={
        choice node,
        draw=black!70!black,
        very thick,
        fill=black!8,
        text=black!70!black,
        font=\scriptsize\bfseries
    },
    neutral edge/.style={
        draw=black!30,
        thin
    },
    selected edge/.style={
        draw=black!70!black,
        very thick
    },
    arrow/.style={
        ->,
        semithick,
        draw=black!50
    },
    selected arrow/.style={
        ->,
        very thick,
        draw=black!70!black
    },
    stage/.style={
        draw=black!45,
        rounded corners=3pt,
        align=center,
        text width=0.72\columnwidth,
        inner sep=5pt
    },
    selected stage/.style={
        stage,
        draw=black!70!black,
        very thick
    },
    token/.style={
        draw=black!45,
        fill=black!3,
        rounded corners=2pt,
        minimum width=11mm,
        minimum height=6.5mm,
        inner sep=2pt,
        font=\ttfamily\footnotesize
    },
    annotation/.style={
        font=\scriptsize,
        text=black!60,
        align=center
    }
]
\node[
    font=\footnotesize\bfseries,
    text=black!70
] (title) at (0,0.65) {Recursive \(k\)-means hierarchy};
\node[root node] (root) at (0,0) {\(d\)};
\tenchildren{depth1}{root}{-0.90}{1}
\tenchildren{depth2}{depth1-1}{-2.15}{8}
\tenchildren{depth3}{depth2-8}{-3.40}{5}
\node[
    stage,
    text width=34mm
] (leaf) at (0,-4.40) {
    Terminal cluster \(C_{185}\)\\[-0.5mm]
    \(\lvert C_{185}\rvert \leq 100\)
};
\draw[arrow] (depth3-5.south) -- (leaf.north);
\node[
    stage,
    below=5mm of leaf
] (prefix) {
    Recursive cluster choices\\[1mm]
    \(1 \mid 8 \mid 5\)\\[1mm]
    Semantic prefix: \texttt{185}
};
\draw[arrow] (leaf) -- (prefix);
\node[
    stage,
    below=4mm of prefix
] (suffix) {
    Local index within \(C_{185}\)\\[1mm]
    \texttt{10}
};
\draw[arrow] (prefix) -- (suffix);
\node[
    stage,
    below=4mm of suffix
] (construction) {
    Semantic identifier assignment\\[1mm]
    \(d \mapsto \texttt{185}\mathbin{\Vert}\texttt{10}
      = \texttt{18510}\)
};
\draw[arrow] (suffix) -- (construction);
\node[
    stage,
    below=4mm of construction
] (tokens) {
    T5 SentencePiece tokens\\[2mm]
    \begin{tikzpicture}[baseline=-0.6ex]
        \node[token] (t1) {\spmark 18};
        \node[token, right=1.5pt of t1] (t2) {510};
        \node[token, right=1.5pt of t2] (t3) {</s>};
    \end{tikzpicture}
};
\draw[arrow] (construction) -- (tokens);
\node[
    stage,
    below=4mm of tokens
] (ids) {
    T5 vocabulary IDs\\[2mm]
    \begin{tikzpicture}[baseline=-0.6ex]
        \node[token] (i1) {507};
        \node[token, right=1.5pt of i1] (i2) {25926};
        \node[token, right=1.5pt of i2] (i3) {1};
    \end{tikzpicture}
};
\draw[arrow] (tokens) -- (ids);
\end{tikzpicture}
\caption{
Construction and tokenization of the semantic identifier \texttt{18510}. Recursive $k$-means routes document $d$ through local= clusters $1$, $8$, and $5$, producing the semantic prefix= \texttt{185}. Once the selected cluster contains 100 or less documents, $d$ receives the leaf-local index \texttt{10}. T5 generates the resulting identifier using the SentencePiece tokens corresponding to \texttt{\_18} and \texttt{510}, followed by EOS.
}
\label{fig:semantic-identifier-tokenization}
\end{figure}

%% file: Appendices/naive-identifier.tex
%

\begin{figure}[t!]
\centering
\begin{tikzpicture}[
    font=\small,
    >=Latex,
    stage/.style={
        draw=black!45,
        rounded corners=3pt,
        align=center,
        text width=0.72\columnwidth,
        inner sep=5pt
    },
    compact stage/.style={
        draw=black!45,
        rounded corners=3pt,
        align=center,
        minimum width=34mm,
        inner sep=5pt
    },
    token/.style={
        draw=black!45,
        fill=black!3,
        rounded corners=2pt,
        minimum width=11mm,
        minimum height=6.5mm,
        inner sep=2pt,
        font=\ttfamily\footnotesize
    },
    number/.style={
        circle,
        draw=black!40,
        minimum size=6mm,
        inner sep=0pt,
        font=\scriptsize
    },
    selected number/.style={
        number,
        draw=black!70,
        very thick,
        fill=black!8,
        font=\scriptsize\bfseries
    },
    arrow/.style={
        ->,
        semithick,
        draw=black!50
    },
    annotation/.style={
        font=\scriptsize,
        text=black!60,
        align=center
    }
]
\node[
    font=\footnotesize\bfseries,
    text=black!70
] (title) {Random identifier assignment};
\node[
    compact stage,
    below=5mm of title
] (document) {
    Document \(d\)
};
\node[
    stage,
    below=4mm of document
] (space) {
    Identifier space\\[1mm]
    \(\{0,1,\ldots,b-1\}\), with \(b\geq\lvert\mathcal D\rvert\)
};
\draw[arrow] (document) -- (space);
\node[
    stage,
    below=4mm of space
] (shuffle) {
    Random permutation with fixed seed\\[2mm]
    \begin{tikzpicture}[baseline=-0.6ex]
        \node[number] (n0) {0};
        \node[number, right=1.5pt of n0] (n1) {1};
        \node[font=\footnotesize, right=2pt of n1] (dots1) {\(\cdots\)};
        \node[selected number, right=2pt of dots1] (chosen) {141913};
        \node[font=\footnotesize, right=2pt of chosen] (dots2) {\(\cdots\)};
        \node[number, right=2pt of dots2] (last) {\(b-1\)};
    \end{tikzpicture}
};
\draw[arrow] (space) -- (shuffle);
\node[
    stage,
    below=4mm of shuffle
] (assignment) {
    Unique random assignment\\[1mm]
    \(d\mapsto\texttt{141913}\)
};
\draw[arrow] (shuffle) -- (assignment);
\node[
    stage,
    below=4mm of assignment
] (identifier) {
    Decimal identifier\\[1mm]
    \(\texttt{141913}\)
};
\draw[arrow] (assignment) -- (identifier);
\node[
    stage,
    below=4mm of identifier
] (tokens) {
    T5 SentencePiece tokens\\[2mm]
    \begin{tikzpicture}[baseline=-0.6ex]
        \node[token] (t1) {\spmark 14};
        \node[token, right=1.5pt of t1] (t2) {19};
        \node[token, right=1.5pt of t2] (t3) {13};
        \node[token, right=1.5pt of t3] (t4) {</s>};
    \end{tikzpicture}
};
\draw[arrow] (identifier) -- (tokens);
\node[
    stage,
    below=4mm of tokens
] (ids) {
    T5 vocabulary IDs\\[2mm]
    \begin{tikzpicture}[baseline=-0.6ex]
        \node[token] (i1) {968};
        \node[token, right=1.5pt of i1] (i2) {2294};
        \node[token, right=1.5pt of i2] (i3) {2368};
        \node[token, right=1.5pt of i3] (i4) {1};
    \end{tikzpicture}
};
\draw[arrow] (tokens) -- (ids);
\end{tikzpicture}
\caption{Construction and tokenization of the naive identifier \texttt{141913}. The identifier space $\{0,\ldots,b-1\}$ is randomly permuted using a fixed seed, and each document receives a distinct integer from the identifier space. T5 tokenizes the resulting decimal string as \texttt{\spmark 14}, \texttt{19}, and \texttt{13}, followed by end-of-sequence, with vocabulary IDs $968$, $2294$, $2368$, and $1$.}
\label{fig:naive-identifier-tokenization}
\end{figure}
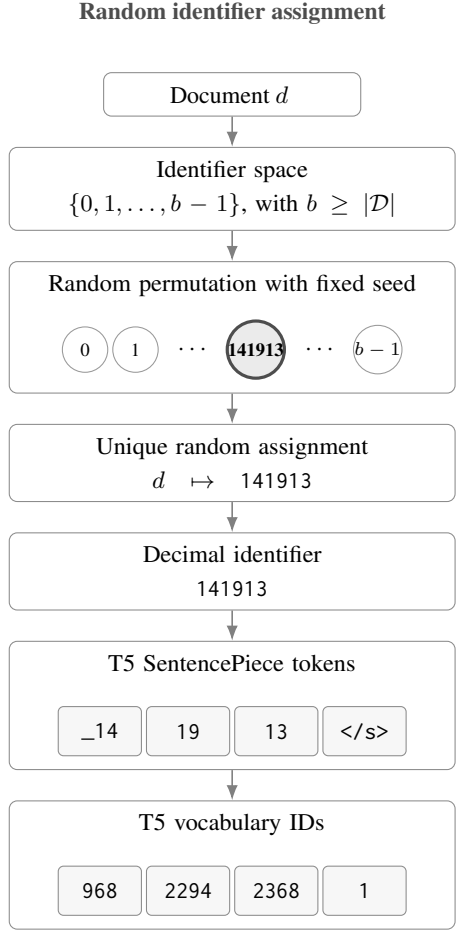

%% file: Appendices/grouping-keys.tex
\section{Document Deduplication Diagnostics}
\label{app:grouping-keys}

\subsection{Comparison Operators and Collection Sizes}
\label{app:grouping-key-counts}

Section~\ref{sec:nq-document-grouping} introduces the four principal
comparison operators $\doceq$ used for document deduplication.
Here we provide their exact implementation details and
additional variants used to isolate individual normalization choices.

\paragraph{Principal comparison operators.}
We define the different comparison operators by describing what strings extracted
from documents they compare.
$\doceq_{\textsc{Text4K}}$ uses the lowercased first \(4{,}000\) Unicode code points
of the serialized document text. $\doceq_{\textsc{URL}}$ uses the complete
revision-specific Wikipedia URL, including its \texttt{oldid}.
$\doceq_{\textsc{Title}}$ follows NCI's split-dependent construction: it uses the
first HTML \htag{h1} element for training records and
\texttt{document\_title} for development records, followed by NCI's
BERT-tokenize-and-decode normalization. The key is computed before
document-text serialization. $\doceq_{\textsc{PageID}}$ uses the recovered persistent
Wikipedia page identifier, with the title-based fallback described in
Section~\ref{sec:nq-document-grouping} for unresolved revisions.

\paragraph{Diagnostic variants.}
To isolate the effect of lowercasing, $\doceq_{\textsc{Text4K-Cased}}$ preserves case,
compared to $\doceq_{\textsc{Text4K}}$ that relies on lowercased version of the same
\(4{,}000\)-code-point prefix. We similarly compare the title extracted
from the Wikipedia URL before and after lowercasing
($\doceq_{\textsc{URLTitle-Cased}}$ and $\doceq_{\textsc{URLTitle}}$), and the surface text of
the first HTML \htag{h1} element before and after lowercasing
($\doceq_{\textsc{H1-Cased}}$ and $\doceq_{\textsc{H1}}$).
Finally, $\doceq_{\textsc{NciH1}}$ applies
NCI's BERT-tokenize-and-decode normalization to the lowercased
\htag{h1} text.

For operators that discard the original HTML
structure, title-derived strings are computed beforehand and retained
alongside the serialized document.

\begin{table}[t]
\centering
\small
\begin{tabular*}{\columnwidth}{@{\extracolsep{\fill}}lrrr@{}}
\toprule
Grouping key & Train & Dev & Union \\
\midrule
Input records
& 307,373
& 7,830
& 315,203 \\
\midrule
\multicolumn{4}{@{}l}{\emph{Text-derived keys}} \\
\addlinespace[2pt]
\textsc{Text4K-Cased}
& 198,918
& 7,280
& 202,970 \\
\textsc{Text4K}
& 198,202
& 7,271
& 202,221 \\
\midrule
\multicolumn{4}{@{}l}{\emph{URL-derived key}} \\
\addlinespace[2pt]
\textsc{URL}
& 226,180
& 7,369
& 231,695 \\
\midrule
\multicolumn{4}{@{}l}{\emph{Title-derived keys}} \\
\addlinespace[2pt]
\textsc{URLTitle-Cased}
& 108,071
& 6,930
& 109,712 \\
\textsc{URLTitle}
& 108,015
& 6,930
& 109,654 \\
\textsc{H1-Cased}
& 108,998
& 6,935
& 110,660 \\
\textsc{H1}
& 108,036
& 6,930
& 109,676 \\
\textsc{NciH1}
& 108,026
& 6,930
& 109,666 \\
\textsc{Title}
& 108,026
& 6,930
& 109,739 \\
\midrule
\multicolumn{4}{@{}l}{\emph{Page-identity key}} \\
\addlinespace[2pt]
\textsc{PageID}
& 107,593
& 6,926
& 109,229 \\
\bottomrule
\end{tabular*}
\caption{Distinct NQ targets produced by different deduplication comparison operators $\doceq$. Train and Dev apply grouping independently within each split. Union applies grouping jointly and gives the indexed collection size. The principal operators are \textsc{Text4K}, \textsc{URL}, \textsc{Title}, and \textsc{PageID}.}
\label{tab:dedup-counts}
\end{table}

\input{Appendices/pageid-resolution}
\input{Appendices/fragmentation-collission-examples}

%% file: Appendices/pageid-resolution.tex
\subsection{PageID Resolution}
\label{app:pageid-resolution}

Natural Questions URLs identify source revisions through an
\texttt{oldid}, but do not directly provide the corresponding Wikipedia
PageID. We recover this mapping from the English Wikipedia
stub-meta-history dump dated December 20, 2018.\footnote{\url{https://archive.org/download/enwiki-20181220/enwiki-20181220-stub-meta-history.xml.gz}}

Of the 231,695 distinct revision identifiers appearing in NQ, 231,589
resolve to a PageID, corresponding to a $99.95\%$ resolution rate.
The remaining 106 identifiers cover 156 NQ records. We retain these
records using a fallback key derived from the URL title and mark them as
unresolved in the grouping audit.

Recovered PageIDs are used only to determine grouping. They do not imply
that revision-specific texts are identical or determine which revision
represents the resulting grouped document.

%% file: Appendices/fragmentation-collission-examples.tex
\subsection{Revision-Specific Content Variation under a Shared Title}
\label{app:shared-title-revision-variation}

Records sharing a title may nevertheless contain substantially different revision-specific text. Title-based deduplication necessarily removes this revision-level distinction, regardless of whether the revisions belong to one persistent PageID or to distinct pages. The examples in Table~\ref{tab:shared-title-revision-variation} are therefore not included in our fragmentation or collision counts, which require resolved PageID evidence.
They illustrate information that is lost when revision records are grouped by title alone.

\begin{table*}[t]
\centering
\small
\begin{tabularx}{\textwidth}{@{}p{0.18\textwidth}YY@{}}
\toprule
Shared title & Revision A & Revision B \\
\midrule
\textit{John Wick}
& \texttt{oldid=804434627}\newline
  \emph{This article is about the 2014 film\ldots}
& \texttt{oldid=843537108}\newline
  \emph{This article is about the action film franchise\ldots} \\
\addlinespace[3pt]
\textit{I Can See Your Voice}
& \texttt{oldid=805035333}\newline
  \emph{South Korean music show by Mnet\ldots}
& \texttt{oldid=826663643}\newline
  \emph{international mystery music game show franchise\ldots} \\
\bottomrule
\end{tabularx}
\caption{Examples of NQ records sharing a title but containing revision-specific text with different apparent scopes. These examples establish content variation but do not determine whether each pair shares a PageID.}
\label{tab:shared-title-revision-variation}
\end{table*}

The \textit{John Wick} records distinguish the 2014 film from the broader franchise, while the \textit{I Can See Your Voice} records distinguish the original South Korean program from the international franchise. These cases show that title equality does not guarantee identical or narrowly equivalent document text. Whether such revisions should form one retrieval document depends on whether the intended retrieval unit is a persistent Wikipedia page or a revision-specific textual snapshot.

\subsection{Rendered-Text Variation under a Fixed Revision}
\label{app:fixed-revision-variation}

A fixed $(\text{title},\texttt{oldid})$ pair does not always determine one unique \texttt{document\_text} string in NQ. We observe 231,695 distinct revision URLs but 241,247 distinct flattened document texts. Inspection of representative cases suggests that some differences arise from Wikipedia template expansion during rendering.

\begin{table*}[t]
\centering
\small
\begin{tabularx}{\textwidth}{@{}p{0.26\textwidth}XX@{}}
\toprule
Article and revision & Variant A & Variant B \\
\midrule
\textit{We Are the World}\newline
\texttt{oldid=815185445}
& \emph{raised over \$63 million (equivalent to \$141 million today)\ldots}
& \emph{raised over \$63 million (equivalent to \$138 million today)\ldots} \\
\addlinespace[3pt]
\textit{List of Billboard Hot 100 number-one singles of 2008}\newline
\texttt{oldid=787133819}
& \emph{By decade: 1930s, 1940s, 1950--1958, 1958--1969, 1970s,\ldots}
& \emph{By decade: 1940s, 1950--1958, 1958--1969, 1970s, 1980s,\ldots} \\
\bottomrule
\end{tabularx}
\caption{Rendered-text variation among NQ records with the same Wikipedia revision identifier.}
\label{tab:fixed-revision-variation}
\end{table*}

In the first example in Table~\ref{tab:fixed-revision-variation}, the article contains the template \texttt{\{\{Inflation|US|63|1986\}\}}, whose rendered present-day value differs across records. The second contains \texttt{\{\{USNumber1s\}\}}, which generates a navigation list whose contents also vary. These examples show that an \texttt{oldid} identifies the underlying Wikipedia revision but does not necessarily determine a unique rendered text in Natural Questions. A revision URL therefore provides a stable source-revision key, but not a one-to-one mapping to the flattened document representation distributed in NQ.

%% file: Appendices/htmlnorm-details.tex
\section{\textsc{HtmlNorm} Reconstruction and Serialization}
\label{app:htmlnorm-details}

\textsc{HtmlNorm} converts the full NQ representation into the textual document sequence used for indexing. The pipeline first recovers surface-faithful text from byte-aligned HTML spans, then serializes structured elements such as lists and tables, and finally applies Wikipedia-specific cleanup rules.

\input{Appendices/html-aware-reconstruction}
\input{Appendices/htmlnorm-serialization-details}

%% file: Appendices/html-aware-reconstruction.tex
\subsection{Byte-Aligned Span Recovery}
\label{app:byte-span-recovery}

Each full NQ record provides the rendered page in \texttt{document\_html} and a sequence of \texttt{document\_tokens}. Each token contains a string, a flag indicating whether it represents an HTML element, and byte offsets into \texttt{document\_html}. The stored token strings are not always surface-faithful: they may normalize punctuation or otherwise differ from the corresponding spans in the rendered HTML. We therefore recover the surface form of each non-HTML token from its byte-aligned HTML span rather than using its stored token string.

Let $H$ denote the UTF-8 byte sequence of \texttt{document\_html}, and let $t_1,\ldots,t_n$ be the non-HTML tokens in document order. Token $t_i$ has byte interval $[s_i,e_i)$, from which we recover
\[
x_i = \operatorname{norm}\!(H[s_i:e_i]),
\]
where $\operatorname{norm}$ decodes the UTF-8 span, resolves HTML entities, and applies Unicode NFKC normalization. This restores surface forms such as quotation marks, brackets, and punctuation that may differ in the stored NQ token strings.

The byte gap between consecutive textual tokens is
\[
g_i = H[e_i:s_{i+1}].
\]
We use the gap and the intervening HTML structure to determine whether the recovered spans should be concatenated, separated by a space, or separated by a line break:
\[
R =
x_1\,\sigma(g_1)\,x_2\,\sigma(g_2)\cdots
\sigma(g_{n-1})\,x_n,
\]
where
\[
\sigma(g)=
\left\{
\begin{array}{@{}ll@{}}
\epsilon,
    & \text{if $g$ is empty},\\[3pt]

\texttt{\char32},
    &
    \begin{array}[t]{@{}l@{}}
    \text{if $g$ contains visible}\\
    \text{inter-token whitespace},
    \end{array}\\[6pt]

\texttt{\textbackslash n},
    &
    \begin{array}[t]{@{}l@{}}
    \text{if $g$ crosses a block-level}\\
    \text{HTML boundary},
    \end{array}\\[6pt]

\epsilon,
    & \text{otherwise}.
\end{array}
\right.
\]
HTML markup alone therefore does not introduce spaces, while visible whitespace and block structure are preserved. This contrasts with the simplified NQ representation, which inserts a space between every pair of stored token strings irrespective of their original HTML spacing.

\subsection{Serialization of Structured Elements}
\label{app:structured-serialization}

\subsubsection{List Formatting}
\label{app:list-formatting}

HTML lists are serialized into line-oriented text so that their item structure remains visible after removing the markup.

\begin{itemize}
    \item Unordered lists (\htag{ul}) use a hyphen marker for each item:
\begin{verbatim}
- item
  - nested item
\end{verbatim}

    \item Ordered lists (\htag{ol}) retain explicit numbering and respect the HTML \texttt{start} attribute:
\begin{verbatim}
1. first item
2. second item
\end{verbatim}

    \item Nested lists are represented by indentation relative to their parent item;

    \item Definition lists (\htag{dl}, \htag{dt}, and \htag{dd}) are serialized as term--definition pairs:
\begin{verbatim}
term: definition
\end{verbatim}

    \item A definition without a preceding term is emitted as a standalone line;

    \item Empty list items and items containing only structural markup are discarded.
\end{itemize}
These rules preserve distinctions that are lost when the NQ token strings are simply space-joined. For example,
\begin{verbatim}
<ol>
  <li>France</li>
  <li>Japan</li>
</ol>
\end{verbatim}
is serialized as
\begin{verbatim}
1. France
2. Japan
\end{verbatim}
rather than \texttt{France Japan}.

\subsubsection{Table Formatting}
\label{app:table-formatting}

HTML tables vary substantially in purpose and structure, so \textsc{htmlnorm} selects a textual serialization according to the detected table layout.

\begin{itemize}
    \item Two-column key--value tables are serialized as attribute--value pairs:
\begin{verbatim}
Born: July 1, 1900
Occupation: Physicist
\end{verbatim}

    \item Tables with meaningful column headers are serialized row by row, with the headers propagated into each row:
\begin{verbatim}
Country: France ; Capital: Paris
Country: Japan ; Capital: Tokyo
\end{verbatim}

    \item Single-column tables are serialized as one line per non-empty row;

    \item Matrix-like tables without a usable header structure are serialized by joining the non-empty cells of each row:
\begin{verbatim}
value A ; value B ; value C
\end{verbatim}

    \item Empty tables and tables containing only structural or interface content are discarded.
\end{itemize}
This serialization favors locally interpretable textual facts over a flat concatenation of cells. Introductory infoboxes are handled separately by the Wikipedia-specific cleanup described below.

%% file: Appendices/htmlnorm-serialization-details.tex
\subsection{Wikipedia-Specific Cleanup}
\label{app:wikipedia-cleanup}

We remove elements that reflect the Wikipedia interface or article-navigation structure rather than the page content used for retrieval. In particular:

\begin{itemize}
    \item interface artifacts such as \texttt{[edit]} and \texttt{[hide]} are removed;

    \item the introductory infobox table is discarded;

    \item processing stops at footer-like sections whose normalized heading matches one of:
    \begin{quote}
    \emph{See also}, \emph{References}, \emph{Further reading},
    \emph{Bibliography}, \emph{External links}, \emph{Notes},
    \emph{Footnotes}, \emph{Citations}, \emph{Sources},
    \emph{Works cited}, \emph{Contents}, or
    \emph{Notes and references}.
    \end{quote}
\end{itemize}

The removal of these sections prevents navigation links, citation lists, and interface text from consuming the fixed model-visible indexing prefix. Other lists and tables occurring in the article body are retained using the serialization rules in Appendices~\ref{app:list-formatting} and~\ref{app:table-formatting}.

\subsection{Pronunciation Cleanup}
\label{app:pronunciation-cleanup}

Wikipedia introductions frequently place pronunciation information immediately after the page title. These parenthetical spans may contain International Phonetic Alphabet symbols or other characters that are outside the T5 SentencePiece vocabulary and are therefore mapped to \unk. Because documents are truncated to a short token prefix during indexing, such spans can consume several model-visible positions while contributing little retrieval content.

We apply a simple heuristic for cleanup. We inspect only the first parenthesized span beginning within the first 256 characters of the reconstructed document. The span is tokenized with the same T5 tokenizer used by the retrieval model. If its tokenization contains at least one \unk{} token, the entire parenthetical span is removed; otherwise, it is retained unchanged.

Restricting the procedure to the first parenthetical span targets the conventional position of Wikipedia pronunciation annotations and avoids removing later parenthetical content throughout the document. The heuristic is nevertheless tokenizer-dependent and may occasionally remove an introductory parenthetical that is not a pronunciation; we therefore report it explicitly as part of the \textsc{htmlnorm} construction.

\subsection{Document Serialization Examples}
\label{app:serialization-examples}

Figure~\ref{fig:nq-text-representation-example} compares document serialization methods.

\begin{figure*}[t!]
\centering

\begin{minipage}[t]{0.315\textwidth}
\vspace{0pt}
\centering
\textbf{\textsc{simplified}}
\vspace{2pt}

\begin{lstlisting}[
    basicstyle=\ttfamily\scriptsize,
    breaklines=true,
    breakatwhitespace=false,
    breakindent=0pt,
    postbreak={},
    columns=fullflexible,
    keepspaces=true,
    showstringspaces=false,
    inputencoding=utf8,
    frame=single,
    aboveskip=0pt,
    belowskip=0pt
]
List of woods - wikipedia List of woods Jump to : navigation , search This is a list of woods , in particular those most commonly used in the timber and lumber trade . Contents ( hide ) 1 Softwoods ( coniferous ) 2 Hardwoods ( angiosperms ) 3 Hardwoods ( monocotyledons ) 4 See also 5 References 6 External links Softwoods ( coniferous ) ( edit ) Araucaria Hoop pine ( Araucaria cunninghamii ) Monkey puzzle tree ( Araucaria araucana ) Parana pine ( Araucaria angustifolia ) Cedar ( Cedrus ) Celery - top pine ( Phyllocladus aspleniifolius ) Cypress ( Chamaecyparis , Cupressus , Taxodium ) Arizona cypress ( Cupressus arizonica ) Bald cypress , southern cypress ( Taxodium distichum ) Alerce ( Fitzroya cupressoides ) Hinoki cypress ( Chamaecyparis obtusa ) Lawson 's cypress ( Chamaecyparis lawsoniana ) Mediterranean cypress ( Cupressus sempervirens ) Douglas - fir ( Pseudotsuga menziesii ) Coast Douglas - fir ( Pseudotsuga menziesii var . menziesii ) Rocky Mountain Douglas - fir ( Pseudotsuga menziesii var . glauca ) European yew ( Taxus baccata ) Fir ( Abies ) Balsam fir ( Abies balsamea ) Silver fir ( Abies alba ) Noble fir ( Abies procera ) Pacific silver fir ( Abies amabilis ) Hemlock ( Tsuga ) Eastern hemlock ( Tsuga canadensis ) Mountain hemlock ( Tsuga mertensiana ) Western hemlock ( Tsuga heterophylla ) Huon pine , Macquarie pine ( Lagarostrobos franklinii ) Kauri ( New Zealand ) ( Agathis australis ) Queensland kauri ( Australia ) ( Agathis robusta ) Japanese nutmeg - yew , kaya ( Torreya nucifera ) Larch ( Larix ) European larch ( Larix decidua ) Japanese larch ( Larix ... [truncated]
\end{lstlisting}
\end{minipage}
\hfill
\begin{minipage}[t]{0.315\textwidth}
\vspace{0pt}
\centering
\textbf{\textsc{simplenorm}}
\vspace{2pt}

\begin{lstlisting}[
    basicstyle=\ttfamily\scriptsize,
    breaklines=true,
    breakatwhitespace=false,
    breakindent=0pt,
    postbreak={},
    columns=fullflexible,
    keepspaces=true,
    showstringspaces=false,
    inputencoding=utf8,
    frame=single,
    aboveskip=0pt,
    belowskip=0pt
]
 List of woods  This is a list of woods , in particular those most commonly used in the timber and lumber trade . Contents ( hide ) 1 Softwoods ( coniferous ) 2 Hardwoods ( angiosperms ) 3 Hardwoods ( monocotyledons ) 4 See also 5 References 6 External links Softwoods ( coniferous ) ( edit ) Araucaria Hoop pine ( Araucaria cunninghamii ) Monkey puzzle tree ( Araucaria araucana ) Parana pine ( Araucaria angustifolia ) Cedar ( Cedrus ) Celery - top pine ( Phyllocladus aspleniifolius ) Cypress ( Chamaecyparis , Cupressus , Taxodium ) Arizona cypress ( Cupressus arizonica ) Bald cypress , southern cypress ( Taxodium distichum ) Alerce ( Fitzroya cupressoides ) Hinoki cypress ( Chamaecyparis obtusa ) Lawson 's cypress ( Chamaecyparis lawsoniana ) Mediterranean cypress ( Cupressus sempervirens ) Douglas - fir ( Pseudotsuga menziesii ) Coast Douglas - fir ( Pseudotsuga menziesii var . menziesii ) Rocky Mountain Douglas - fir ( Pseudotsuga menziesii var . glauca ) European yew ( Taxus baccata ) Fir ( Abies ) Balsam fir ( Abies balsamea ) Silver fir ( Abies alba ) Noble fir ( Abies procera ) Pacific silver fir ( Abies amabilis ) Hemlock ( Tsuga ) Eastern hemlock ( Tsuga canadensis ) Mountain hemlock ( Tsuga mertensiana ) Western hemlock ( Tsuga heterophylla ) Huon pine , Macquarie pine ( Lagarostrobos franklinii ) Kauri ( New Zealand ) ( Agathis australis ) Queensland kauri ( Australia ) ( Agathis robusta ) Japanese nutmeg - yew , kaya ( Torreya nucifera ) Larch ( Larix ) European larch ( Larix decidua ) Japanese larch ( Larix kaempferi ) Tamarack ( Larix laricina ) Western larch ... [truncated]
\end{lstlisting}
\end{minipage}
\hfill
\begin{minipage}[t]{0.315\textwidth}
\vspace{0pt}
\centering
\textbf{\textsc{htmlnorm}}
\vspace{2pt}

\begin{lstlisting}[
    basicstyle=\ttfamily\scriptsize,
    breaklines=true,
    breakatwhitespace=false,
    breakindent=0pt,
    postbreak={},
    columns=fullflexible,
    keepspaces=true,
    showstringspaces=false,
    inputencoding=utf8,
    frame=single,
    aboveskip=0pt,
    belowskip=0pt
]
List of woods
This is a list of woods, in particular those most commonly used in the timber and lumber trade.

Softwoods (coniferous)

- Araucaria
  - Hoop pine (Araucaria cunninghamii)
  - Monkey puzzle tree (Araucaria araucana)
  - Parana pine (Araucaria angustifolia)
- Cedar (Cedrus)
- Celery-top pine (Phyllocladus aspleniifolius)
- Cypress (Chamaecyparis, Cupressus, Taxodium)
  - Arizona cypress (Cupressus arizonica)
  - Bald cypress, southern cypress (Taxodium distichum)
  - Alerce (Fitzroya cupressoides)
  - Hinoki cypress (Chamaecyparis obtusa)
  - Lawson's cypress (Chamaecyparis lawsoniana)
  - Mediterranean cypress (Cupressus sempervirens)
- Douglas-fir (Pseudotsuga menziesii)
  - Coast Douglas-fir (Pseudotsuga menziesii var. menziesii)
  - Rocky Mountain Douglas-fir (Pseudotsuga menziesii var. glauca)
- European yew (Taxus baccata)
- Fir (Abies)
  - Balsam fir (Abies balsamea)
  - Silver fir (Abies alba)
  - Noble fir (Abies procera)
  - Pacific silver fir (Abies amabilis)
- Hemlock (Tsuga)
  - Eastern hemlock (Tsuga canadensis)
  - Mountain hemlock (Tsuga mertensiana)
  - Western hemlock (Tsuga heterophylla)
- Huon pine, Macquarie pine (Lagarostrobos franklinii)
- Kauri (New Zealand) (Agathis australis)
- Queensland kauri (Australia) (Agathis robusta)
- Japanese nutmeg-yew, kaya (Torreya nucifera)
- Larch (Larix)
  - European larch (Larix decidua)
  - Japanese larch (Larix kaempferi)
  - Tamarack (Larix laricina)
  - Western larch (Larix occidentalis)
- Pine (Pinus)
  - European black pine (Pinus nigra)
  - Jack pine (Pinus banksiana)
  - Lodgepole pine (Pinus ... [truncated]
\end{lstlisting}
\end{minipage}

\caption{Three textual representations constructed from the same NQ page snapshot. \textsc{simplified} removes the stored HTML tags but retains the flattened token sequence and page-interface content. \textsc{simplenorm} selects the title, abstract, and subsequent article content from that sequence. \textsc{htmlnorm} reconstructs the original surface text from the byte-aligned HTML and preserves readable structure. Excerpts are truncated to the same display budget.}
\label{fig:nq-text-representation-example}
\end{figure*}

%% file: Appendices/natural-questions.tex
\section{Natural Questions}

This appendix describes the Natural Questions (NQ) data formats and illustrates how Wikipedia page content is represented after preprocessing.

\subsection{NQ Formats}
\label{app:nq-formats}

\input{Tables/nq-format-fields}

\begin{figure*}[t]
    \begin{minipage}[t]{0.45\textwidth}
\begin{lstlisting}[
          basicstyle=\ttfamily\scriptsize,
          breaklines=true,
          breakatwhitespace=false,
          breakindent=0pt,
          postbreak={},
          columns=fullflexible,
          keepspaces=true,
          inputencoding=utf8
        ]
{ 
  "example_id": 6915606477668963399,
  "question_text": "what do the 3 dots mean in math",
  "question_tokens": [
    "what", "do", "the", "3", "dots", "mean", "in", "math"
  ],
  "document_url": "https://en.wikipedia.org//w/index.php?
                   title=Therefore_sign&amp;oldid
                   =815234923",
    
  "document_title": "Therefore sign",
  "document_html": "<!DOCTYPE html>\n<HTML class=\"client-
                    js ve-not-available\" lang=\"en\" dir
                    =\"ltr\"><HEAD>\n\n<TITLE>Therefore 
                    sign - Wikipedia</TITLE>\n\n\...",
  "document_tokens": [
    {
      "token": "Therefore"
      "html_token": false,
      "start_byte": 92,
      "end_byte": 101,
    },
    ...
  ],
  "long_answer_candidates": [
    {
      "top_level": true,
      "start_token": 14,
      "end_token": 808,
      "start_byte": 41427,
      "end_byte": 66428
    },
    ...
  ],
  "annotations": [
    {
      "annotation_id": 13591449469826568799,
      "long_answer": {
        "candidate_index": 92,
        "start_token": 808,
        "end_token": 925,
        "start_byte": 66429,
        "end_byte": 67824
      },
      "short_answers": [
        {
          "start_token": 816,
          "end_token": 837,
          "start_byte": 66588,
          "end_byte": 66817
        }
      ],
      "yes_no_answer": "NONE"
    },
    ...
  ]
}
\end{lstlisting}
    \end{minipage}\hfill%
    \begin{minipage}[t]{0.45\textwidth}
\begin{lstlisting}[
            basicstyle=\ttfamily\scriptsize,
            breaklines=true,
            breakatwhitespace=false,
            breakindent=0pt,
            postbreak={},
            columns=fullflexible,
            keepspaces=true,
            inputencoding=utf8
        ]
{ 
  "example_id": 6915606477668963399,
  "question_text": "what do the 3 dots mean in math",
  "document_url": "https://en.wikipedia.org//w/index.php?
                   title=Therefore_sign&amp;oldid
                   =815234923",
    
  "document_text": "Therefore sign - wikipedia <H1> 
                    Therefore sign </H1> Jump to : 
                    navigation , search <Table> <Tr> <
                    Th_colspan=\"2\"> ...",
  "long_answer_candidates": [
    { 
      "top_level": true,
      "start_token": 14,
      "end_token": 808
    },
    ...
  ],
  "annotations": [
    {
      "annotation_id": 13591449469826568799,
      "long_answer": {
        "candidate_index": 92,
        "start_token": 808,
        "end_token": 925
      },
      "short_answers": [
        {
          "start_token": 816,
          "end_token": 837,
        }
      ],
      "yes_no_answer": "NONE"
    },
  ],
  ...
}
\end{lstlisting}
    \end{minipage}
    \caption{Example NQ page record in the full format (left) and simplified format (right).}
    \label{fig:nq-format-example}
\end{figure*}

Natural Questions is distributed in two formats, referred to as \emph{full} and \emph{simplified}. In addition to metadata, a Wikipedia snapshot may be represented through the following fields:
\begin{itemize}
    \item \texttt{document\_html} contains the original HTML snapshot;
    \item \texttt{document\_tokens} contains the article content selected by the NQ preprocessing pipeline and segmented into text and HTML-tag tokens;
    \item \texttt{document\_text} serializes \texttt{document\_tokens} by inserting spaces between stored tokens and therefore does not preserve the original spacing;
    \item \texttt{document\_title} contains the Wikipedia page title, extracted from the HTML \htag{title} element after removing the trailing \texttt{ - Wikipedia} suffix.
\end{itemize}
Table~\ref{tab:nq-format-fields} shows which fields are available in each format, and Figure~\ref{fig:nq-format-example} provides an example of both.

The training data is distributed in both full and simplified formats, whereas the development set is distributed only in the full format.

Unfortunately, the processing pipeline used to build the \texttt{document\_tokens} field is unavailable.
The content is tokenized into words and HTML tags, including byte offsets that retain a link to the original full HTML page.
However, the tokenized strings may already contain substitutions, and these substitutions can alter the resulting T5 tokenization.
We analyze these substitutions in Appendix~\ref{app:nq-token-substitutions}.

\subsection{Character Substitutions in the NQ Preprocessing Pipeline}
\label{app:nq-token-substitutions}

To characterize information lost by the NQ preprocessing pipeline, we compare each visible token string in \texttt{document\_tokens} with its byte-aligned span in \texttt{document\_html}.
We exclude HTML-tag tokens, compare case-insensitively, and treat HTML-entity-equivalent forms as matches.
Across the 7,830 development records, \(1.88\)M of \(54.13\)M compared tokens differ from their original spans, corresponding to \(3.47\%\) of the tokens and 22,333 distinct replacement pairs.
The ten most frequent pairs, shown in Table~\ref{tab:nq-token-substitutions}, account for \(94.6\%\) of all observed mismatches.

\begin{table}[t]
\centering
\small
\setlength{\tabcolsep}{3pt}
\begin{tabular*}{\columnwidth}{@{\extracolsep{\fill}}llccrr@{}}
\toprule
&
&
\multicolumn{2}{c}{Contains \(\unk\)}
&
&
\\
\cmidrule(lr){3-4}
HTML
& Stored token
& HTML
& Stored
& Count
& Records \\
\midrule
\texttt{\char34}
& \texttt{\char96\char96}
& \textendash
& \(\checkmark\)
& 564,605
& 7,830 \\
\texttt{\char34}
& \texttt{\char39\char39}
& \textendash
& \textendash
& 555,899
& 7,830 \\
\textendash
& \texttt{--}
& \textendash
& \textendash
& 289,496
& 6,902 \\
\texttt{]}
& \texttt{)}
& \textendash
& \textendash
& 145,151
& 7,791 \\
\texttt{[}
& \texttt{(}
& \textendash
& \textendash
& 144,996
& 7,791 \\
\textemdash
& \texttt{--}
& \textendash
& \textendash
& 30,361
& 3,970 \\
\texttt{\{}
& \texttt{(}
& \(\checkmark\)
& \textendash
& 18,845
& 351 \\
\texttt{\}}
& \texttt{)}
& \(\checkmark\)
& \textendash
& 18,804
& 366 \\
\textquoteright\texttt{s}
& \texttt{\char39 s}
& \textendash
& \textendash
& 6,965
& 2,217 \\
\textquotedblright
& \texttt{\char39\char39}
& \textendash
& \textendash
& 4,528
& 556 \\
\bottomrule
\end{tabular*}
\caption{Most frequent substitutions between visible strings in
\texttt{document\_tokens} and their byte-aligned spans in
\texttt{document\_html} on the NQ development set.
HTML entities are shown after unescaping.
Under the two \(\unk\) columns, \(\checkmark\) indicates that the
corresponding string produces an unknown token with the T5 tokenizer,
while \textendash\ indicates that it does not.
The same quotation-mark character is mapped to opening
\texttt{\char96\char96} or closing \texttt{\char39\char39} tokens
according to context.
Count gives the number of token occurrences, and Records gives the number
of affected NQ examples.
These ten replacement pairs account for \(94.6\%\) of all observed
mismatches.}
\label{tab:nq-token-substitutions}
\end{table}

Most mismatches arise from systematic normalization of quotation marks, dashes, and brackets rather than isolated corruption.
These substitutions always change the surface form and may change the T5 token sequence even when neither form produces \(\unk\).
Their effect on vocabulary coverage is not uniform: for example, the stored opening-quotation token introduces an unknown token, whereas replacing braces with parentheses removes one.
Representations built from \texttt{document\_tokens} as \textsc{SimpleNorm} inherit these transformations.
In \textsc{HtmlNorm}, we instead recover the corresponding byte spans from \texttt{document\_html}.

\subsection{Whitespace Artifacts and T5 Tokenization}
\label{app:nq-tokenization-comparison}

The NQ preprocessing pipeline segments article content into text and HTML-tag tokens. Reconstructing a textual sequence by inserting spaces between these stored tokens can introduce whitespace artifacts and punctuation changes, which in turn alter the subsequent T5 tokenization.

Consider the subsequence \enquote{\texttt{matt" lanter (born april 1,}} from the second example in Table~\ref{tab:ipa-tokenization-example}.
The NQ preprocessing pipeline transforms it into \enquote{\texttt{matt '' lanter ( born april 1 ,}}.
As such, if we compare the result of the T5 tokenization on (1)~the output of \textsc{NciNorm} versus (2)~the output of \textsc{HtmlNorm}, we obtain (underlining the main differences and using \texttt{\textvisiblespace} for whitespace characters):\footnote{Note that the difference in the first token is due to the previous unknown token in the full example.}
\begin{enumerate}
\item
\enquote{\tok{\textvisiblespace{}mat}},
\enquote{\tok{'t}},
\underline{\enquote{\tok{\textvisiblespace{}}},
\enquote{\tok{'}},
\enquote{\tok{'}}},
\enquote{\tok{\textvisiblespace{}}},
\enquote{\tok{lant}},
\enquote{\tok{er}},
\enquote{\tok{\textvisiblespace{}(}},
\enquote{\tok{\textvisiblespace{}born}},
\enquote{\tok{\textvisiblespace{}}},
\enquote{\tok{a}},
\enquote{\tok{pri}},
\enquote{\tok{l}},
\underline{\enquote{\tok{\textvisiblespace{}1}},
\enquote{\tok{\textvisiblespace{}}},
\enquote{\tok{,}}}.

\item
\enquote{\tok{mat}},
\enquote{\tok{t}},
\underline{\enquote{\tok{"}}},
\enquote{\tok{\textvisiblespace{}}},
\enquote{\tok{lant}},
\enquote{\tok{er}},
\enquote{\tok{\textvisiblespace{}(}},
\enquote{\tok{born}},
\enquote{\tok{\textvisiblespace{}}},
\enquote{\tok{a}},
\enquote{\tok{pri}},
\enquote{\tok{l}},
\underline{\enquote{\tok{\textvisiblespace{}1,}}}.
\end{enumerate}

Compared with \textsc{NciNorm}, \textsc{HtmlNorm} avoids additional tokens introduced by whitespace artifacts, which would otherwise consume positions in the fixed-length input budget.

%% file: Tables/nq-format-fields.tex
\begin{table}[t]
\centering
\small
\begin{tabular*}{\columnwidth}{@{\extracolsep{\fill}}lcc@{}}
\toprule
Field or information & Full & Simplified \\
\midrule
\texttt{example\_id}              & \checkmark & \checkmark \\
\texttt{question\_text}           & \checkmark & \checkmark \\
\texttt{question\_tokens}         & \checkmark & \textendash \\
\texttt{document\_url}\textsuperscript{\(\dagger\)}
                                  & \checkmark & \checkmark \\
\texttt{document\_title}\textsuperscript{\(\dagger\)}
                                  & \checkmark & \textendash \\
\texttt{document\_html}\textsuperscript{\(\dagger\)}
                                  & \checkmark & \textendash \\
\texttt{document\_tokens}\textsuperscript{\(\dagger\)}
                                  & \checkmark & \textendash \\
\texttt{document\_text}\textsuperscript{\(\dagger\)}
                                  & \textendash & \checkmark \\
\midrule
\multicolumn{3}{@{}l}{\emph{Alignment information}} \\
\addlinespace[2pt]
Document-token HTML byte offsets\textsuperscript{\(\dagger\)}
                                  & \checkmark & \textendash \\
Long-answer token offsets         & \checkmark & \checkmark \\
Long-answer byte offsets          & \checkmark & \textendash \\
Short-answer token offsets        & \checkmark & \checkmark \\
Short-answer byte offsets         & \checkmark & \textendash \\
\bottomrule
\end{tabular*}
\caption{Availability of the principal fields and alignment information in the full and simplified NQ formats. Entries marked with \(\dagger\) are used by at least one document-construction or grouping variant considered in this work. The byte offsets associated with \texttt{document\_tokens} link each stored token to its original span in \texttt{document\_html}.}
\label{tab:nq-format-fields}
\end{table}

%% file: Appendices/indexing-retrieval-ratio.tex
\section{Indexing-to-Retrieval Ratio}
\label{app:indexing-retrieval-ratio}

The expectations in Eq.\ \eqref{eq:obj} are estimated with different numbers of samples.
In this appendix, an indexing-to-retrieval ratio of $a:b$ means than for $a$ samples to approximate the indexing term, there are $b$ samples to approximate the retrieval term, no matter the minibatch size.

DSI reports sampling indexing and retrieval examples at an indexing-to-retrieval ratio of $32{:}1$ \citep{tay2022dsi}. Our default instead follows their natural frequency in \textsc{NQ320K-Title/NciNorm}: each document contributes one indexing example and each labeled query contributes one retrieval example, yielding an approximately $1{:}2.8$ ratio. We evaluate this choice with T5-Base while holding the identifier-specific hyperparameters, effective batch size, and training budget fixed. Indexing-heavy ratios are obtained by repeating document examples, while $0{:}1$ removes indexing supervision entirely.

\input{Tables/indexing-retrieval-ratio}

Indexing supervision is necessary for all three identifier types: removing it substantially reduces retrieval effectiveness. Increasing the indexing weight beyond the natural ratio also generally degrades performance. A balanced $1{:}1$ ratio yields the highest MRR@10 for atomic and naive identifiers, but only marginally exceeds the natural ratio and improves mainly Hits@10 rather than Hits@1. The natural ratio remains best for semantic identifiers.

Invalidity follows a different pattern. For semantic identifiers, Invalid@40 decreases steadily as the ratio becomes more indexing-heavy, even while retrieval effectiveness deteriorates; naive identifiers show the same tendency through $32{:}1$. We therefore retain the natural ratio as a common setting: it avoids identifier-specific tuning while remaining close to the best retrieval effectiveness for all three identifier types.

%% file: Tables/indexing-retrieval-ratio.tex
\begin{table*}[t]
\centering
\small
\setlength{\tabcolsep}{3pt}
\begin{tabular*}{\textwidth}{@{\extracolsep{\fill}}lrrrrrrrrrrrr@{}}
\toprule
&
\multicolumn{4}{c}{\textbf{Atomic}} &
\multicolumn{4}{c}{\textbf{Naive}} &
\multicolumn{4}{c}{\textbf{Semantic}} \\
\cmidrule(lr){2-5}
\cmidrule(lr){6-9}
\cmidrule(lr){10-13}
Ratio
& H@1 & H@10 & MRR@10 & Inv.
& H@1 & H@10 & MRR@10 & Inv.
& H@1 & H@10 & MRR@10 & Inv. \\
\midrule
$0{:}1$
& 51.35 & 65.52 & 56.35 & \textendash
& 46.39 & 56.82 & 50.03 & 38.5
& 46.23 & 56.95 & 49.92 & 26.0 \\
Natural ($\approx1{:}2.8$)
& \textbf{62.80} & 81.56 & 69.42 & \textendash
& \textbf{55.29} & 69.03 & 60.07 & 35.5
& \textbf{54.39} & \textbf{68.16} & \textbf{59.10} & 24.4 \\
$1{:}1$
& 62.50 & \textbf{83.15} & \textbf{69.76} & \textendash
& 55.25 & \textbf{69.53} & \textbf{60.25} & 33.4
& 53.74 & 67.41 & 58.33 & 22.8 \\
$4{:}1$
& 59.32 & 81.02 & 66.84 & \textendash
& 53.83 & 68.61 & 59.01 & 31.7
& 52.54 & 66.97 & 57.40 & 17.3 \\
$8{:}1$
& 56.96 & 77.98 & 64.15 & \textendash
& 53.49 & 68.98 & 58.95 & 31.3
& 51.17 & 66.07 & 56.21 & 13.2 \\
$16{:}1$
& 54.56 & 74.90 & 61.44 & \textendash
& 52.53 & 69.30 & 58.49 & 29.8
& 49.97 & 65.94 & 55.35 & 11.1 \\
$32{:}1$
& 54.39 & 72.01 & 60.51 & \textendash
& 48.88 & 66.73 & 55.07 & 28.7
& 46.74 & 64.11 & 52.40 & 7.5 \\
$64{:}1$
& 52.53 & 72.30 & 59.29 & \textendash
& 44.13 & 63.83 & 50.88 & 31.7
& 42.41 & 61.35 & 48.87 & 5.7 \\
\bottomrule
\end{tabular*}
\caption{Sensitivity to the indexing-to-retrieval ratio under \textsc{NQ320K-Title/NciNorm}. Retrieval metrics and Invalid@40 are development percentages at the checkpoint selected by MRR@10. The natural ratio uses each indexed document and labeled query once per repetition of the training data. Invalidity is not applicable to atomic identifiers and is denoted by \textendash. Bold marks the best retrieval result for each identifier type and metric.}
\label{tab:indexing-retrieval-ratio}
\end{table*}

%% file: Appendices/hyperparameters-search.tex
\section{Hyperparameter Search Results}
\label{app:hyperparameters-search}

Table~\ref{tab:hparams-controlled-full} reports the complete batch-128 sweep used to select the configurations for the experiments.

For atomic identifiers, we retain $\ell=64$, a learning rate of $5\times10^{-4}$, and 10K warmup steps; naive and semantic identifiers use $\ell=32$, a learning rate of $10^{-3}$, and no warmup.
These settings are reused throughout our experiments.

\begin{table}[t!]
\centering
\small
\setlength{\tabcolsep}{2.5pt}
\begin{tabular*}{\columnwidth}{@{\extracolsep{\fill}}lrrrrrrr@{}}
\toprule
ID & $\ell$ & LR & WU & H@1 & H@10 & M@10 & Inv@40 \\
\midrule
\multicolumn{8}{@{}l}{\emph{Atomic}}\\
Atomic & 32 & $1\mathrm{e}{-3}$ & 0    & 61.4 & 79.1 & 67.7 & \textendash \\
Atomic & 32 & $1\mathrm{e}{-3}$ & 10K  & 61.8 & 81.1 & 68.5 & \textendash \\
Atomic & 32 & $1\mathrm{e}{-3}$ & 100K & 59.0 & 80.2 & 66.4 & \textendash \\
Atomic & 32 & $5\mathrm{e}{-4}$ & 0    & 62.1 & 79.3 & 68.1 & \textendash \\
Atomic & 32 & $5\mathrm{e}{-4}$ & 10K  & 62.3 & 80.5 & 68.6 & \textendash \\
Atomic & 32 & $5\mathrm{e}{-4}$ & 100K & 60.3 & 79.9 & 67.3 & \textendash \\
Atomic & 64 & $1\mathrm{e}{-3}$ & 0    & 61.4 & 79.0 & 67.6 & \textendash \\
Atomic & 64 & $1\mathrm{e}{-3}$ & 10K  & 61.2 & 80.3 & 67.8 & \textendash \\
Atomic & 64 & $1\mathrm{e}{-3}$ & 100K & 59.7 & 79.7 & 66.6 & \textendash \\
Atomic & 64 & $5\mathrm{e}{-4}$ & 0    & 62.0 & 79.4 & 68.1 & \textendash \\
\textbf{Atomic}
& \textbf{64}
& $\mathbf{5\mathrm{e}{-4}}$
& \textbf{10K}
& 62.0
& 81.1
& \textbf{68.7}
& \textendash \\
Atomic & 64 & $5\mathrm{e}{-4}$ & 100K & 61.0 & 80.4 & 67.6 & \textendash \\
\addlinespace[2pt]
\multicolumn{8}{@{}l}{\emph{Naive}}\\
\textbf{Naive}
& \textbf{32}
& $\mathbf{1\mathrm{e}{-3}}$
& \textbf{0}
& 55.3
& 69.4
& \textbf{60.1}
& 31.9 \\
Naive & 32 & $1\mathrm{e}{-3}$ & 10K  & 55.1 & 69.1 & 60.1 & 35.8 \\
Naive & 32 & $1\mathrm{e}{-3}$ & 100K & 54.6 & 67.9 & 59.1 & 37.2 \\
Naive & 32 & $5\mathrm{e}{-4}$ & 0    & 54.7 & 68.4 & 59.5 & 36.3 \\
Naive & 32 & $5\mathrm{e}{-4}$ & 10K  & 54.7 & 68.6 & 59.5 & 37.2 \\
Naive & 32 & $5\mathrm{e}{-4}$ & 100K & 54.4 & 67.6 & 59.2 & 38.4 \\
Naive & 64 & $1\mathrm{e}{-3}$ & 0    & 55.5 & 69.0 & 60.2 & 33.6 \\
Naive & 64 & $1\mathrm{e}{-3}$ & 10K  & 55.0 & 68.6 & 59.7 & 36.0 \\
Naive & 64 & $1\mathrm{e}{-3}$ & 100K & 54.5 & 67.9 & 59.2 & 36.4 \\
Naive & 64 & $5\mathrm{e}{-4}$ & 0    & 54.2 & 68.0 & 59.0 & 36.5 \\
Naive & 64 & $5\mathrm{e}{-4}$ & 10K  & 54.3 & 67.9 & 59.1 & 37.4 \\
Naive & 64 & $5\mathrm{e}{-4}$ & 100K & 53.8 & 67.6 & 58.6 & 38.6 \\
\addlinespace[2pt]
\multicolumn{8}{@{}l}{\emph{Semantic}}\\
\textbf{Semantic}
& \textbf{32}
& $\mathbf{1\mathrm{e}{-3}}$
& \textbf{0}
& 55.2
& 68.4
& \textbf{59.7}
& 22.9 \\
Semantic & 32 & $1\mathrm{e}{-3}$ & 10K  & 54.5 & 67.6 & 58.9 & 22.3 \\
Semantic & 32 & $1\mathrm{e}{-3}$ & 100K & 54.1 & 67.5 & 58.6 & 24.6 \\
Semantic & 32 & $5\mathrm{e}{-4}$ & 0    & 54.0 & 68.2 & 58.8 & 21.9 \\
Semantic & 32 & $5\mathrm{e}{-4}$ & 10K  & 53.8 & 67.4 & 58.4 & 24.6 \\
Semantic & 32 & $5\mathrm{e}{-4}$ & 100K & 53.8 & 67.2 & 58.3 & 23.4 \\
Semantic & 64 & $1\mathrm{e}{-3}$ & 0    & 54.7 & 68.4 & 59.3 & 22.1 \\
Semantic & 64 & $1\mathrm{e}{-3}$ & 10K  & 54.1 & 67.6 & 58.6 & 21.8 \\
Semantic & 64 & $1\mathrm{e}{-3}$ & 100K & 53.6 & 66.8 & 58.1 & 24.4 \\
Semantic & 64 & $5\mathrm{e}{-4}$ & 0    & 53.2 & 66.8 & 57.8 & 22.4 \\
Semantic & 64 & $5\mathrm{e}{-4}$ & 10K  & 53.0 & 67.1 & 57.8 & 22.5 \\
Semantic & 64 & $5\mathrm{e}{-4}$ & 100K & 52.6 & 66.7 & 57.3 & 23.0 \\
\bottomrule
\end{tabular*}
\caption{Complete batch-128 hyperparameter sweep on \textsc{NQ320K-Title/NciNorm}. Metrics are reported as percentages; M@10 denotes MRR@10. Each row reports the checkpoint selected by development MRR@10. Bold hyperparameters identify the configurations retained for the experiments; for naive identifiers, $\ell=32$ is retained as a near-tie with the grid maximum at $\ell=64$. Dashes indicate that Invalid@40 is not applicable to atomic identifiers, which are valid by construction.}
\label{tab:hparams-controlled-full}
\end{table}

%% file: Appendices/reproduction-settings.tex
\section{Implementation Details}
\label{app:reproduction-settings}

This appendix documents implementation details relevant to reproducing \textsc{ReDSI}. Unless stated otherwise, experiments use the settings described in Section~\ref{sec:experimental-setup}. The remaining baselines in Table~\ref{tab:published-dsi-nq} are included only for context.

\subsection{Software and Hardware}
\label{app:software-hardware}

\textsc{ReDSI} is implemented with Hugging Face Transformers version \texttt{4.44.1} and trained on NVIDIA V100 GPUs with 16\,GB of memory using FP16 mixed precision. We use T5~1.0 checkpoints and the tokenizer distributed with each checkpoint.

Atomic identifiers are implemented by extending the decoder output vocabulary with one document-specific token per indexed document. Naive and semantic identifiers use the original T5 vocabulary and are generated autoregressively. Models are optimized with Adafactor using the identifier-specific learning rates and warmup durations selected in Appendix~\ref{app:hyperparameters-search}, with no learning-rate decay after warmup.

\begin{table}[t!]
\centering
\scriptsize
\begin{tabular*}{\columnwidth}{@{\extracolsep{\fill}}lll@{}}
\toprule
Setting
& \citet{pradeep2023does}
& \textsc{ReDSI} \\
\midrule
Collection
& NCI-derived, ${\sim}109$K
& \textsc{Title/NciNorm} \\
Backbone
& T5~1.1-Base
& T5~1.0-Base \\
Supervision
& FirstP + labeled queries
& FirstP + labeled queries \\
Input length
& 64
& 64 atomic; 32 sequential \\
Effective batch size
& 512
& 128 \\
Sequence packing
& Not reported
& Disabled \\
Checkpoint selection
& Not reported
& Development MRR@10 \\
Invalid outputs
& Not reported
& Post-filtered \\
\bottomrule
\end{tabular*}
\caption{Reported and implemented settings for the closest comparison with \citet{pradeep2023does}. Differences and unreported settings prevent an exact reproduction.}
\label{tab:pradeep-settings}
\end{table}

\subsection{Framework Differences}
\label{app:framework-differences}

Our Transformers input pipeline does not use sequence packing:\footnote{\url{https://huggingface.co/docs/trl/v1.12.0/en/reducing_memory_usage\#packing}} each batch element corresponds to one padded indexing or retrieval example. Consequently, the effective batch size counts original training examples. T5X supports packing multiple examples into shared sequences, but \citet{pradeep2023does} do not report whether packing was enabled. If it was, their nominal batch size may contain more original examples than an unpacked Transformers batch of the same size.

Additional differences may remain in token-loss normalization, auxiliary loss terms, identifier construction, beam-search implementation, and checkpoint selection. Because these settings are not fully documented, we do not attribute the remaining differences in effectiveness to any single factor. Sensitivity to the indexing-to-retrieval ratio is reported separately in Appendix~\ref{app:indexing-retrieval-ratio}.

%% file: Appendices/nq320k-construction-variability.tex
\section{Run-to-Run Variability of NQ320K Construction Effects}
\label{app:nq320k-construction-variability}

We repeat every T5-Efficient-Tiny identifier--construction configuration with seeds 42--46, reusing the main-experiment hyperparameters and selecting checkpoints by development MRR@10, for a total of 60 runs. Tables~\ref{tab:representation-grouping-seeds} and~\ref{tab:representation-grouping-effects} report the resulting variability and paired construction effects.

\begin{table}[t]
\centering
\small
\begin{tabular*}{\columnwidth}{@{\extracolsep{\fill}}lrr@{}}
\toprule
Identifier
& \textsc{NciNorm}
& \textsc{HtmlNorm} \\
\midrule
\multicolumn{3}{c}{\textsc{Title}} \\
\midrule
Atomic
& $62.09 \pm 0.10$
& $62.54 \pm 0.06$ \\
Naive
& $46.53 \pm 0.40$
& $46.86 \pm 0.40$ \\
Semantic
& $47.12 \pm 0.21$
& $48.09 \pm 0.06$ \\
\midrule
\multicolumn{3}{c}{\textsc{PageID}} \\
\midrule
Atomic
& $63.25 \pm 0.10$
& $63.49 \pm 0.06$ \\
Naive
& $47.52 \pm 0.36$
& $47.76 \pm 0.50$ \\
Semantic
& $48.72 \pm 0.10$
& $48.77 \pm 0.23$ \\
\bottomrule
\end{tabular*}
\caption{T5-Efficient-Tiny MRR@10 across five seeds for each identifier, grouping rule, and document representation. Values are mean percentages with sample standard deviations.}
\label{tab:representation-grouping-seeds}
\end{table}

\begin{table}[t]
\centering
\small
\begin{tabular*}{\columnwidth}{@{\extracolsep{\fill}}lrr@{}}
\toprule
Identifier
& \textsc{HtmlNorm} $-$ \textsc{NciNorm}
& \textsc{PageID} $-$ \textsc{Title} \\
\midrule
Atomic
& $+0.34 \pm 0.04$
& $+1.06 \pm 0.10$ \\
Naive
& $+0.29 \pm 0.42$
& $+0.94 \pm 0.24$ \\
Semantic
& $+0.51 \pm 0.23$
& $+1.14 \pm 0.22$ \\
\midrule
Global
& $\mathbf{+0.38 \pm 0.18}$
& $\mathbf{+1.05 \pm 0.10}$ \\
\bottomrule
\end{tabular*}
\caption{Seed-paired effects of document representation and grouping for T5-Efficient-Tiny. Values are changes in MRR@10 percentage points, reported as mean $\pm$ sample standard deviation across five seeds. Identifier-level effects are averaged over the other construction factor; the global result macro-averages identifier types within each seed.}
\label{tab:representation-grouping-effects}
\end{table}

%% file: Appendices/invalid-40.tex
\section{Invalidity across Construction Variants and Model Scales}
\label{app:invalid-40}

Table~\ref{tab:invalid-40} reports the complete Invalid@40 results underlying the model-scaling analysis in Section~\ref{sec:identifier-analysis}.

\input{Tables/invalid-40}

%% file: Tables/invalid-40.tex
\begin{table*}[t!]
\centering
\small
\begin{tabular*}{\textwidth}{@{\extracolsep{\fill}}lrrrrrrrr@{}}
\toprule
&
\multicolumn{4}{c}{\textbf{\textsc{Title}}} &
\multicolumn{4}{c}{\textbf{\textsc{PageID}}} \\
\cmidrule(lr){2-5}
\cmidrule(lr){6-9}
&
\multicolumn{2}{c}{\textsc{NciNorm}} &
\multicolumn{2}{c}{\textsc{HtmlNorm}} &
\multicolumn{2}{c}{\textsc{NciNorm}} &
\multicolumn{2}{c}{\textsc{HtmlNorm}} \\
\cmidrule(lr){2-3}
\cmidrule(lr){4-5}
\cmidrule(lr){6-7}
\cmidrule(lr){8-9}
Model
& Naive & Semantic
& Naive & Semantic
& Naive & Semantic
& Naive & Semantic \\
\midrule
T5-Base
& 31.9 & 22.9
& 35.7 & 20.6
& 33.0 & 23.4
& 35.6 & 22.2 \\

T5-Small
& 31.0 & 20.3
& 31.2 & 19.4
& 30.6 & 18.8
& 29.3 & 19.6 \\

T5-Eff.-Tiny
& 37.5 & 8.7
& 36.8 & 8.6
& 38.2 & 8.5
& 37.4 & 9.1 \\
\bottomrule
\end{tabular*}

\caption{Invalid@40 for sequential identifiers across NQ320K grouping rules, document representations, and model scales. Values are percentages measured on the unconstrained 40-best beam before post-filtering. Atomic identifiers are omitted because they are valid by construction.}
\label{tab:invalid-40}
\end{table*}

%% file: Appendices/additional-decoding-results.tex
\section{Decoding Procedures and Additional Results}
\label{app:additional-decoding-results}

Let $Y=\{\docid(d) \mid d\in\docs\}$ denote the set of valid identifiers. Sequential identifiers are decoded with beam size 40. We distinguish raw unconstrained decoding, post-filtering, and trie-constrained decoding.

\subsection{Decoding Procedures}
\label{app:decoding-procedures}

\paragraph{Raw unconstrained decoding.}
Standard beam search operates over the complete T5 vocabulary without restricting generated prefixes. For a query $\vq$, it returns a ranked beam
\[
\mathcal{B}(\vq)=\bigl(\vy^{(1)},\ldots,\vy^{(40)}\bigr),
\]
ordered by the model's length-normalized sequence score. A hypothesis is invalid when $\vy^{(i)}\notin Y$. Raw decoding retains invalid hypotheses in the ranking, where they may displace valid identifiers from the first $k$ positions.

\paragraph{Post-filtered decoding.}
Post-filtering applies the same unconstrained search and removes completed hypotheses that do not belong to $\mathcal{Y}$:
\[
\mathcal{B}_{\mathrm{valid}}(\vq)
=
\bigl\{y\in\mathcal{B}(\vq)\mid y\in Y\bigr\}.
\]
The scores and relative order of the retained hypotheses are unchanged. Because filtering occurs after search, the resulting ranking may contain fewer than 40 documents and cannot recover valid hypotheses that fell outside the original beam.

We report \emph{Invalid@40}, the mean fraction of unconstrained hypotheses that are removed:
\[
\operatorname{Invalid@40}
=
\frac{1}{\card{Q}}
\sum_{\vq\in Q}
\frac{
\card*{\{y\in\mathcal{B}(\vq)\mid y\notin Y\}}
}{40}.
\]

\paragraph{Trie-constrained decoding.}
Trie-constrained beam search restricts generation to prefixes of identifiers in $Y$ \citep{cao2021entityretrieval}. For a generated prefix $\bm u$, the permitted next tokens are
\[
\operatorname{Next}(u)\!=\!\{v\!\in\!\voc\!\mid\!uv\text{ is a prefix of some }y\in Y\}.
\]
Tokens outside $\operatorname{Next}(u)$ are masked before beam expansion, and the end-of-sequence token is permitted only when the completed sequence belongs to $Y$. Constrained decoding therefore guarantees valid completed hypotheses and reallocates capacity otherwise spent on invalid prefixes. It does not guarantee relevance: a valid identifier may still correspond to a non-relevant document.

\subsection{Identifier-Space Budget Sensitivity}
\label{app:decoding-budget}

For naive identifiers, we vary the assignment-space budget from $b=\card{\docs}$ to $b=32\times\card{\docs}$, including DSI's $b=320\mathrm{K}\approx2.9\cdot\card{\docs}$. Larger budgets make assignments sparser but also change identifier tokenization, length, and prefix sharing. We therefore treat the sweep as a sensitivity analysis rather than as an isolated intervention on invalidity. All other settings, including the identifier-assignment seed, remain fixed.

\begin{figure}[t]
\centering
\resizebox{0.7\columnwidth}{!}{%
  \input{pgf/naive_budget_decoding.pgf}%
}
\caption{MRR@10 and Invalid@40 for naive identifiers across identifier-space budgets on \textsc{NQ320K-Title}/\textsc{NciNorm}. Dotted, solid, and dashed lines denote raw, post-filtered, and trie-constrained decoding, respectively; the dash-dotted line denotes Invalid@40. DSI corresponds to $b=320\mathrm{K}$.}
\label{fig:budget-decoding}
\end{figure}
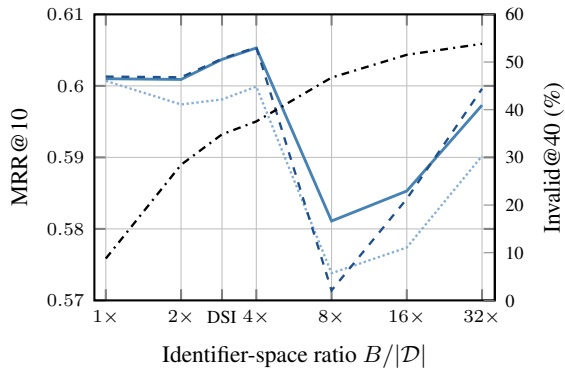

Figure~\ref{fig:budget-decoding} compares retrieval effectiveness under raw, post-filtered, and trie-constrained decoding, together with the Invalid@40 rate of the underlying unconstrained beam.

As shown in Figure~\ref{fig:budget-decoding}, Invalid@40 increases from $8.8\%$ at $1\times$ to $53.8\%$ at $32\times$, whereas MRR@10 varies
non-monotonically. The widening gap between raw and post-filtered decoding at larger budgets indicates that filtering becomes more useful as invalid
hypotheses occupy more of the beam, although it does not change the overall budget trend. Trie-constrained decoding closely matches post-filtering
through $4\times$ and has mixed effects at larger budgets, changing MRR@10 by between $-0.97$ and $+0.23$ percentage points. It therefore guarantees valid outputs but provides no consistent effectiveness advantage over post-filtering in the evaluated settings.

%% file: pgf/naive_budget_decoding.pgf
\begin{tikzpicture}[trim axis left, trim axis right]
\begin{axis}[
    name=effectiveness,
    scale only axis,
    width=0.7\linewidth,
    height=0.5\linewidth,
    xmode=log,
    log basis x=2,
    xmin=0.9,
    xmax=36,
    ymin=0.57,
    ymax=0.61,
    ytick={0.57,0.58,0.59,0.60,0.61},
    xtick={1,2,2.916,4,8,16,32},
    xticklabels={
        \(1\times\),
        \(2\times\),
        DSI,
        \(4\times\),
        \(8\times\),
        \(16\times\),
        \(32\times\)
    },
    xlabel={Identifier-space ratio \(B/|\mathcal{D}|\)},
    ylabel={MRR@10},
    yticklabel style={
        /pgf/number format/fixed,
        /pgf/number format/precision=2
    },
    tick label style={font=\scriptsize},
    label style={font=\small},
    axis line style={line width=0.8pt},
    tick style={line width=0.8pt},
    grid=major,
    clip=true,
]

\addplot[
    naivelight,
    densely dotted,
    line width=0.9pt
]
coordinates {
    (1,     0.6007)
    (2,     0.5974)
    (2.916, 0.5981)
    (4,     0.5999)
    (8,     0.5738)
    (16,    0.5774)
    (32,    0.5902)
};

\addplot[
    naivemid,
    solid,
    line width=1.1pt
]
coordinates {
    (1,     0.6010)
    (2,     0.6009)
    (2.916, 0.6037)
    (4,     0.6053)
    (8,     0.5811)
    (16,    0.5853)
    (32,    0.5973)
};

\addplot[
    naivedark,
    dashed,
    line width=0.9pt
]
coordinates {
    (1,     0.6013)
    (2,     0.6012)
    (2.916, 0.6038)
    (4,     0.6054)
    (8,     0.5714)
    (16,    0.5842)
    (32,    0.5996)
};

\end{axis}

\begin{axis}[
    overlay,
    at={(effectiveness.south west)},
    anchor=south west,
    scale only axis,
    width=0.70\linewidth,
    height=0.50\linewidth,
    xmode=log,
    log basis x=2,
    xmin=0.9,
    xmax=36,
    ymin=0,
    ymax=0.60,
    axis x line=none,
    xtick=\empty,
    xlabel={},
    axis y line*=right,
    ylabel={Invalid@40 (\%)},
    ytick={0,0.1,0.2,0.3,0.4,0.5,0.6},
    yticklabels={0,10,20,30,40,50,60},
    tick label style={font=\scriptsize},
    label style={font=\small},
    line width=0.8pt,
    grid=none,
    clip=true,
]

\addplot[
    black,
    dash dot,
    line width=0.9pt
]
coordinates {
    (1,     0.0878)
    (2,     0.2845)
    (2.916, 0.3483)
    (4,     0.3754)
    (8,     0.4674)
    (16,    0.5152)
    (32,    0.5382)
};

\end{axis}

\end{tikzpicture}